\PassOptionsToPackage{dvipsnames}{xcolor}
\documentclass[twocolumn]{openjournal}

\usepackage{natbib}
\usepackage{colorblind}
\usepackage{graphicx}
\usepackage{xcolor}
\usepackage{amsmath}
\usepackage{array}
\usepackage{tikz}
\usetikzlibrary{calc,fit,positioning,arrows.meta,backgrounds,shapes.geometric}
\definecolor{oiGreen}{HTML}{009E73}
\definecolor{oiOrange}{HTML}{D55E00}
\definecolor{oiBlue}{HTML}{0072B2}
\definecolor{oiGray}{HTML}{4D4D4D}

\newcommand{\teff}{\ensuremath{T_{\rm eff}}}
\newcommand{\logg}{\ensuremath{\log g}}
\newcommand{\feh}{\mbox{[Fe/H]}}
\newcommand{\mgh}{\mbox{[Mg/H]}}
\newcommand{\kms}{\ensuremath{\,{\rm km\,s^{-1}}}}
\newcommand{\dchi}{\ensuremath{\Delta\chi^2}}
\newcommand{\fimp}{\ensuremath{f_{\rm imp}}}
\newcommand{\Gaia}{{\it Gaia}}
\defcitealias{2018MNRAS.476..528E}{EB18}
\defcitealias{2018MNRAS.473.5043E}{EB18a}
\providecommand{\facilities}[1]{\par\noindent\textit{Facilities:} #1\par}
\providecommand{\software}[1]{\par\noindent\textit{Software:} #1\par}
\makeatletter\def\frontmatter@title@below{\vspace*{-9pt}}\makeatother

\begin{document}

\title{Spectroscopic Binary Detection as Agent-Callable Tools:\\
Detecting 40{,}000+ Main-Sequence Binary Candidates from SDSS DR19 APOGEE Spectra}

\author{Serat M. Saad}
\affiliation{Department of Astronomy, The Ohio State University, 140 West 18th Avenue, Columbus, OH 43210, USA}
\email{saad.104@osu.edu}

\author{Yuan-Sen Ting}
\affiliation{Department of Astronomy, The Ohio State University, 140 West 18th Avenue, Columbus, OH 43210, USA}
\affiliation{Center for Cosmology and AstroParticle Physics (CCAPP), The Ohio State University, Columbus, OH 43210, USA}
\affiliation{Max-Planck-Institut f\"ur Astronomie, K\"onigstuhl 17, D-69117 Heidelberg, Germany}
\email{ting.74@osu.edu}


\begin{abstract}
Unresolved binaries are common in spectroscopic surveys, and their blended light biases the parameters inferred for them. Methods to detect them exist, but applying one to a new data release is limited mostly by operating know-how that is rarely written down. We package the double-lined spectroscopic binary decomposition of \citet{2018MNRAS.476..528E} for reuse on APOGEE spectra, with the executable operations as nine tool servers built on the Model Context Protocol (MCP) and the operating decisions as a Skill. Run over the $238{,}205$ DR19 dwarfs under a fixed driver script, the classifier flags 41{,}466 SB2 candidates, about fifteen times the $2{,}645$ identified in DR13, though not at matched purity. The $8.1\%$ control false-positive rate implies that close to $40\%$ are single stars, so we release the sample as a candidate list. The joint visit fit prefers two stars for $46\%$ of the SB2 with multiple visits. Of these, $4{,}149$ have velocity changes above $10~\kms$. This cut is also passed by $7\%$ of flagged single stars. The pipeline velocities identify another $3{,}943$ velocity variables. We compare the eccentricities of the best-sampled binaries and find no significant difference between the close twins ($q>0.95$) and matched non-twins, which excludes an eccentricity excess of the kind measured at wide separations. An ablation of two operating decisions illustrates that a fresh agent recovers a removed decision only when its absence leaves a measurable trace in the fit. We release the tools, the Skill, and the DR19 catalog with its multi-epoch supplement.
\end{abstract}

\keywords{Spectroscopic binary stars, Close binary stars, Astronomy software, Astronomy data analysis, Sky surveys, Multiple stars}

\section{Introduction}\label{sec:intro}

Stars are commonly found not in isolation but as members of multiple systems. The multiplicity fraction depends on primary mass and on birth environment, from the near-unity companion frequency of O and B stars \citep{2007ApJ...670..747K, 2012Sci...337..444S} through the $\sim$40--50\% of solar-type field dwarfs \citep{2010ApJS..190....1R, 2011ApJ...731....8K} to the cataloged hierarchies of the solar neighborhood \citep{2018ApJS..235....6T}. The distributions of mass ratio, period, and eccentricity that go with it are a basic input to stellar astrophysics, recording the fragmentation and dynamical processing of star-forming gas \citep{2013ARA&A..51..269D, 2017ApJS..230...15M, 2023ASPC..534..275O, 1995MNRAS.277.1491K, 2020MNRAS.491.5158T} and the dynamical history of the Galaxy \citep{2016ARA&A..54..529B}. A multi-object spectroscopic survey \citep[e.g.][]{2006AJ....132.1645S, 2012RAA....12.1197C, 2019Msngr.175....3D, 2024MNRAS.530.2688J, 2023A&A...674A..29R} therefore observes a substantial population of unresolved multiples, in which two or more stars fall within a single fiber and their light combines into one composite spectrum.

\subsection{Spectroscopic binaries and their detection}\label{sec:intro-binaries}

Binaries can be identified in several ways. Astrometrically, an unresolved companion displaces the photocenter and raises the \Gaia\ renormalized unit weight error \citep[RUWE;][]{2020MNRAS.496.1922B, 2022MNRAS.513.2437P, 2021A&A...649A...2L} and can be fit as an astrometric orbit \citep{2023A&A...674A...9H, 2023MNRAS.518.2991S, 2019A&A...623A..72K}, while wide, resolved pairs are selected directly from \Gaia\ astrometry \citep{2018MNRAS.480.4884E, 2021MNRAS.506.2269E}; see \citet{2024NewAR..9801694E} for a review of \Gaia's impact on binary-star science. Spectroscopically, radial-velocity variations across epochs give orbits \citep{1994ApJ...420..806Z, 2004A&A...424..727P, 2017ApJ...837...20P}, an approach that has been used for the close-binary population of APOGEE \citep{2020MNRAS.499.1607M, 2021AJ....162..184K, 2022MNRAS.512.2051D}. The present work concerns the third case, the spectral signature of two stars blended in a single exposure.

Unresolved multiplicity is both a science target and a source of systematic error. When a single-star model is fit to a composite spectrum, the inferred effective temperature, surface gravity, and chemical abundances are biased, most of all at intermediate mass ratio where the secondary contributes enough flux to change the lines without producing a second set of lines \citep{2018MNRAS.473.5043E}. Both data-driven \citep[e.g.][]{2015ApJ...808...16N, 2019ApJ...879...69T} and physical \citep[e.g.][]{2016AJ....151..144G} pipelines for deriving these stellar parameters inherit this bias equally, since both fit a single template to the data and have no way to represent a second component.

Recovering that second component from a single epoch is therefore both a correction to the stellar parameters and a measurement of the binary in its own right. Two families of method are used at survey scale. Cross-correlation approaches look for multiple peaks in the correlation of the observed spectrum against a template, and have been applied to Gaia-ESO \citep{2017A&A...608A..95M} and to APOGEE \citep{2021AJ....162..184K}. They target the double-lined spectroscopic binaries (SB2), in which the two sets of absorption lines are separated enough to appear as distinct peaks.

Forward-modeling approaches instead build a two-component composite from a single-star spectral model, fit it to the observed spectrum, and compare the result to the best single-star fit. The decision rests on whether the second component improves the fit, not on whether it produces a countable second peak, so the classical SB2 is the large-separation limit of a wider class of composite spectra. \citet{2018MNRAS.473.5043E} introduced the composite forward model and \citet{2018MNRAS.476..528E} built the DR13 SB2 catalog and the acceptance thresholds we adopt (hereafter \citetalias{2018MNRAS.476..528E}, and \citetalias{2018MNRAS.473.5043E} for the method paper). We adopt the forward-modeling decomposition of \citetalias{2018MNRAS.476..528E} as our worked example, because it is fully specified, has a public implementation,\footnote{\url{https://github.com/kareemelbadry/binspec}} and has a published DR13 catalog to compare against. \citet{2022MNRAS.517..356K} and \citet{2024MNRAS.527..521K} applied their method to LAMOST-MRS and reported $2{,}460$ and $12{,}426$ SB2 candidates, respectively. \citet{2026arXiv260714794K} also identified $2{,}813$ composite spectra in RAVE. The method was developed from a Gaia-ESO study of M11 \citep{2022MNRAS.510.1515K}, which followed \citetalias{2018MNRAS.476..528E}. It uses ab initio models and fits the flux ratio directly, without an isochrone. In \citet{2024MNRAS.527..521K}, this is done through the radius ratio. Candidates are selected using projected rotational velocity and visual inspection, with more emphasis on purity than completeness.

\subsection{Publishing method know-how for agents}\label{sec:intro-agents}

A method like the \citetalias{2018MNRAS.476..528E} decomposition is communicated as a paper and, increasingly, as public code. Neither one passes on the operating experience that makes the method run correctly on a new data set. The main choices are the normalization convention that cancels in the detection statistic, the pixels that must be masked, the signal-to-noise cap, the multi-start strategy for the mass ratio, the calibration of the acceptance thresholds, and the failure modes that give fake detections. These decisions are mostly tacit. They are spread across the source paper, the code, the calibration data, and the authors' judgment, and they are rarely written down as a procedure. A reader with the equations and the public code still has to work them out again, which is the step that limits the reuse of a published method.

Language models are beginning to change how such analyses are done. The capability rests on a line of work in machine learning: prompting a model to reason step by step \citep{2022arXiv220111903W}, interleaving that reasoning with actions \citep[ReAct;][]{2022arXiv221003629Y}, teaching models to call external tools \citep[Toolformer;][]{2023arXiv230204761S}, and combining several such agents \citep{2023arXiv230808155W}, all built on the transformer architecture \citep{2017arXiv170603762V} and on general-purpose models \citep{2020arXiv200514165B, 2022arXiv220302155O, 2021arXiv210807258B, 2023arXiv230308774O}, some of which learn to browse and use tools directly \citep{2021arXiv211209332N}. In the sciences these agents now run chemistry platforms \citep{2023arXiv230405376B, 2023Natur.624..570B} and end-to-end research loops \citep{2024arXiv240806292L}. Beyond their role as literature and reasoning assistants \citep{2024MNRAS.527.1494L, 2025NatSR..1513751D}, large language models (LLMs) are used in astronomy as agents that call scientific tools and iterate under quality control. These tools are already advancing stellar astrophysics and survey analysis \citep{2026OJAp....9.1879T,2024arXiv240920252F, 2024MNRAS.530.1935S}.

Two ingredients make it possible to publish a method for an agent rather than for a human reader: Model Context Protocol (MCP) and Skill. The MCP is an open standard for exposing a capability as a typed tool with declared inputs, outputs, and units, which any agent can call. A Skill is a written record of the operating decisions and their order, given to the agent so that it applies the method the way an expert would. Together, MCP and Skill let us publish a method's operating know-how alongside its equations and code, in a form an agent can re-apply to new data.

In this paper we do this for the \citetalias{2018MNRAS.476..528E} SB2 decomposition, packaging it as MCP servers and a Skill and running it as an agent over APOGEE DR19. Our main result is the DR19 catalog, a main-sequence SB2 sample consistent with \citetalias{2018MNRAS.476..528E}. A multi-epoch extension refits the individual visits, with the joint fit preferring two stars for $46\%$ of the SB2 with multiple visits and $94\%$ of those also identified as double-lined in published surveys, and adds velocity variables and candidates for orbit fitting (Section~\ref{sec:res-multiepoch}). We use the best-sampled systems for a first eccentricity comparison of close twins against non-twins (Section~\ref{sec:res-ecc}). We also ask which of the operating decisions an agent could recover on its own. Removing one decision from the Skill and giving a fresh agent the tools and the ablated Skill separates the know-how that leaves a measurable trace in the fit from the know-how that does not (Section~\ref{sec:res-blind}).

\section{Data}\label{sec:data}

\subsection{APOGEE DR19}\label{sec:data-apogee}

We use the APOGEE arm \citep{2017AJ....154...94M} of the Sloan Digital Sky Survey (SDSS; \citealt{2017AJ....154...28B}) DR19, part of the fifth-generation survey SDSS-V \citep{2017arXiv171103234K}, recorded with the $2.5$-m Sloan Telescope \citep{2006AJ....131.2332G} and the APOGEE spectrograph \citep{2019PASP..131e5001W} and reduced by the Astra~0.6.0 pipeline into \texttt{mwmStar} products \citep{2015AJ....150..173N, 2020ApJS..249....3A, 2023ApJS..267...44A}. Each combined spectrum is sampled on an 8575-pixel logarithmic-wavelength grid spanning 15100.8--16999.8~\AA. We read the raw flux and inverse variance and do not use the survey continuum, since the normalization is re-derived as part of the fit. The spectra are delivered in the rest frame, with the radial velocity applied upstream. We verified that applying any additional velocity shift to the delivered spectra strictly increases the fit $\chi^2$ (degrades the fit), confirming the rest-frame state.

We select main-sequence dwarfs from the DR19 \texttt{the\_payne\_apogee\_star} catalog, whose labels are computed with the method of \citet{2019ApJ...879...69T}, requiring surface gravity $\logg$ between 4 and 5, effective temperature $\teff$ between 4000 and 7000~K, and signal-to-noise ratio above 60. This yields 238{,}205 dwarfs with usable combined spectra, the full sample over which the search is run. The 11{,}424 primaries below the 4200~K floor of the single-star model (Section~\ref{sec:analysis-apogee}; 4.8\% of the sample) are retained because line doubling remains detectable there, but are fit at the floor and carry a larger minimum detectable mass ratio, which we fold into the completeness discussion.

\subsection{Validation samples}\label{sec:data-valid}

Two labeled sets underpin the calibration of the acceptance thresholds (Section~\ref{sec:acceptance}) and the completeness and benchmark reporting: an enlarged benchmark whose positives are {\it real} \citetalias{2018MNRAS.476..528E}-flagged SB2, and an injection-recovery completeness set. Where we inject, we add synthetic secondaries to {\it real} single-star spectra, so that the injected composites carry the same instrumental residuals, telluric remnants, and label imperfections that the classifier must survive on genuine data. Injecting secondaries into model spectra instead would omit exactly the structure that generates false positives.

For the benchmark we take the DR19 spectra of the \citetalias{2018MNRAS.476..528E}-flagged SB2 that survive our quality cuts, $2{,}344$ systems, as positives, and a matched set of $7{,}866$ non-flagged dwarfs spanning the same signal-to-noise range as controls. The false-positive rate is measured on the held-out controls the classifier never trained on, and this labeled set defines the recovery-versus-false-positive curves of Section~\ref{sec:res-bench}, where the adopted operating point is set.

For completeness we draw real single-star primaries across the temperature range, attach a secondary at a grid of mass ratios $q$ and velocity separations $\Delta v$, synthesize the isochrone-tied companion with the catalog single-star model (Section~\ref{sec:analysis-apogee}), sum in flux with the luminosity weighting of Equation~(1), and add the primary's own inverse-variance noise. The recovered fraction in each $(q,\teff)$ bin is the completeness mapped in Figure~\ref{fig:qteff}.

Because the companion comes from the same model the classifier uses, the recovered fractions are an upper bound on the true completeness, since a real binary carries line-list and abundance differences that the model does not reproduce and is harder to recover than a model composite. This is the same reason we calibrate the acceptance thresholds on real held-out controls rather than on model injections (Section~\ref{sec:acceptance}). We use this map for the shape of the sensitivity, how the minimum detectable $q$ increases toward cool primaries, and we do not turn it into an absolute multiplicity fraction.

For external validation we cross-match the confirmed sample against the \citet{2021AJ....162..184K} APOGEE cross-correlation catalog and against \Gaia\ \citep{2016A&A...595A...1G} DR3 \citep{2023A&A...674A...1G} astrometry and non-single-star solutions \citep{2023A&A...674A..34G}, which are not used in the spectroscopic detection. Recovery and false-positive rates are quoted with binomial (Wilson) 68\% intervals given these sample sizes.

\section{Methods}\label{sec:analysis}

\subsection{The single-star model}\label{sec:analysis-apogee}

The classifier accepts a star as a binary when a two-component fit beats the single-star fit. Its load-bearing ingredient is the single-star model, which must predict a normalized spectrum from stellar labels accurately enough that the residual of a genuine single star is consistent with noise. We follow \citetalias{2018MNRAS.476..528E} in building it data-driven, learning the label-to-spectrum map from the survey's own spectra \citep[e.g.][]{2015ApJ...808...16N, 2016arXiv160303040C, 2019ApJ...879...69T, 2019ApJS..245...34X, 2018MNRAS.475.2978F, 2019MNRAS.483.3255L, 2006ApJ...636..804A, 2006MNRAS.370..141R, 2014A&A...569A.111B} rather than synthesizing it from a model atmosphere and a line list \citep{2003IAUS..210P.A20C, 2008A&A...486..951G, 2012MNRAS.427...27B}, calibrated against benchmark stars \citep{2015A&A...582A..49H} and tied where needed to a stellar-evolution grid \citep{2011ApJS..192....3P, 2016ApJS..222....8D}; Section~\ref{sec:disc-instrument} measures what the synthetic route costs.

For APOGEE we use the five-label network of \citet{2019ApJ...879...69T}, a three-layer perceptron that maps the label vector $\theta = (\teff, \logg, \feh, \mgh, v_{\rm macro})$ through two hidden layers of 300 units to a continuum-normalized spectrum $f(\lambda;\theta)$, sampled on the $7{,}214$ continuum pixels of the \citetalias{2018MNRAS.476..528E} normalization and trained with a Huber loss and early stopping. The map runs from labels to spectrum. Fitting a star inverts it, optimizing $\theta$ until $f(\lambda;\theta)$ matches the observed spectrum. This single map is the whole of the single-star model, and the two-component composite of Section~\ref{sec:analysis-method} is built by evaluating it once per component. Its label coverage therefore bounds the search. The network has a $4200$~K temperature floor, and primaries below it are fit at the floor (Section~\ref{sec:data-apogee}), which raises the mass ratio at which their companions stay detectable (Section~\ref{sec:res-catalog}).

Training the network requires (label, spectrum) pairs, and the labels are the weak link. The DR19 pipeline labels are produced by a different code, and that mismatch acts as target noise that both blurs the learned map and generates false positives.

We therefore use the network to clean its own training labels, in three steps. A first network is trained on $\sim$$4{,}800$ curated dwarfs with their pipeline labels. This is the open reproduction network examined in Section~\ref{sec:disc-instrument}. That network then refits, once, the labels of the $7{,}179$ dwarfs of the underlying control sample that a preliminary detection pass with the first network leaves unflagged as candidate binaries. We call these $7{,}179$ stars the refit pool, and split it in half. The catalog network is trained on the refit labels of a random half ($3{,}589$ stars). The other half ($3{,}590$ stars) is never seen in training and provides the held-out controls on which the catalog's false-positive rate is measured (Section~\ref{sec:res-bench}). The $7{,}866$ benchmark controls of Section~\ref{sec:data-valid} come from the same underlying control sample and serve the separate purpose of setting the recovery against false-positive curves, so the three counts refer to the refit pool, its held-out half, and the benchmark set in turn.

The refit is anchored, in that the new labels are seeded from and stay tied to the pipeline labels rather than iterated freely, and this is what makes the (label, spectrum) training pairs self-consistent with the model family. Iterating the refit further degrades the classifier, because the labels drift away from their physical anchoring while the fit residual stays flat. The same architecture is used for every variant compared in Section~\ref{sec:res-bench}: its fidelity to the data, not the network architecture, is what sets the classifier's performance. The training set, the refit labels, and the trained network are all built from public DR19 data and are released with this work.

\subsection{The forward model and the detection statistic}\label{sec:analysis-method}

\begin{figure}
\centering
\includegraphics[width=\columnwidth]{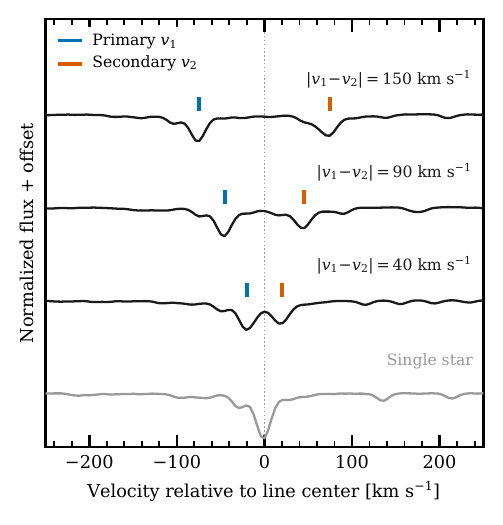}
\caption{How an SB2 imprints line doubling, illustrated with a $q=0.90$ composite model built the same way the classifier builds it (a Sun-like primary combined with a cooler secondary through the isochrone tie), shown around one strong, isolated $H$-band line. A single star gives one absorption core (gray); as the component velocity separation $|v_1-v_2|$ grows, the core splits into a resolved pair at the primary (blue) and secondary (orange) velocities. This velocity-dependent doubling, summed over the many lines in the spectrum, is the signal the two-component model captures.\label{fig:doubling}}
\end{figure}

\begin{figure*}
\centering
\includegraphics[width=\textwidth]{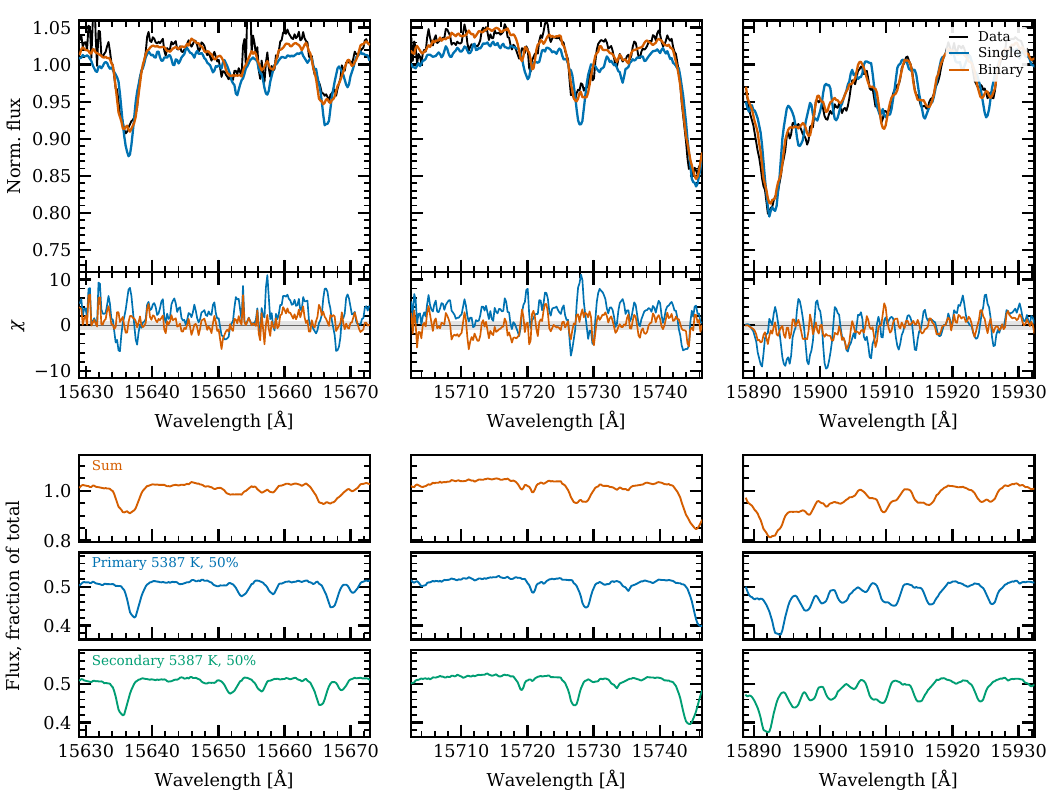}
\caption{A DR19 dwarf SB2 recovered by the agent (\texttt{sdss\_id} 116011140, a near-equal-mass pair), at the three $H$-band windows where the binary model most improves the fit. Top: data with the best single-star (blue) and binary (orange) models. Middle: the $\chi$ residuals; the single-star model swings past the $\pm1\sigma$ band at the doubled line cores the binary model captures. The lower rows show the two-component model in orange, the primary in blue, and the secondary in green. Each is shown in a separate row on the true flux scale. The flux of each component is divided by the continuum of the combined spectrum, so adding the two components gives the model. The horizontal axis of each column is observed wavelength over one of the three windows. Our fit excludes pixels that the DR19 pipeline flags as unusable by setting their inverse variance to effectively zero. In these windows, only the first two pixels of the third window are excluded. Masked pixels are shaded gray. We evaluated the models using the catalog's fitting layer and network, with the labels from the refit used for the released $v_{\rm macro}$. This refit gives $q=1.00$ and $\dchi=34{,}333$. The catalog entry comes from the first census fit, which settled in a shallower minimum with $q=0.91$ and $\dchi=33{,}095$.\label{fig:sb2fit}}
\end{figure*}

The classifier compares two models of the same spectrum: the single-star model above and a two-component composite built from it. The composite model implemented in our servers follows \citetalias{2018MNRAS.476..528E}. Two continuum-normalized single-star spectra are summed with a weight set by each component's band luminosity and shifted by the two component velocities,
\begin{equation}
f_{\rm bin}(\lambda) = \frac{w_1\, f_1(\lambda; v_1) + w_2\, f_2(\lambda; v_2)}{w_1 + w_2},
\end{equation}
where $f_{1,2}=f(\lambda;\theta_{1,2})$ are the two components' single-star spectra from the model of Section~\ref{sec:analysis-apogee}, $v_{1,2}$ their line-of-sight velocities, and the weight $w_i = R_i^2\, B_\lambda(T_{{\rm eff},i})$ combines each component's radius $R_i$ with the mean $H$-band surface brightness $B_\lambda(T_{{\rm eff},i})$ (a Planck factor), turning the surface-area weight $R_i^2$ into a band-luminosity weight. The temperatures, radii, and hence the weights all follow from the primary and a single mass ratio $q$ through a MIST isochrone \citep{2016ApJ...823..102C, 2016ApJS..222....8D} at the primary's metallicity and an assumed common age of 4~Gyr, a representative main-sequence age for the APOGEE dwarf sample, so the flux ratio is fixed by $q$ and not fitted (the isochrone-tied luminosity weighting of Section~\ref{sec:analysis-skill}). The primary mass $M_1$ is located on the 4-Gyr isochrone at the primary's ($\teff$, $\logg$, $\feh$), the secondary mass is $M_2=q\,M_1$, and $T_{{\rm eff},2}$, $L_2$, and hence $R_2$ are read off the same isochrone at $M_2$, which enforces the coeval, equal-composition pair. We therefore tie the abundances rather than fitting them separately, as in \citetalias{2018MNRAS.476..528E}, so both components carry the primary's $\feh$ and $\mgh$. The isochrone is evaluated by direct linear interpolation over the MIST main-sequence grid, which is single-valued in mass, with the metallicity and age snapped to the nearest grid nodes; no neural-net emulator is involved, unlike in the \texttt{binspec} implementation. At $q=1$ and zero velocity separation the composite reduces exactly to the single-star model at the primary's labels, and a nonzero separation still leaves the line doubling that the classifier reads. The composite therefore adds only three parameters over the single-star fit, $q$, $v_1$, and $v_2$.

The velocity shift is what the classifier keys on. Once $|v_1-v_2|$ approaches the instrumental resolution, a single absorption line appears in the composite as a resolved doubled core that splits further as the separation grows (Figure~\ref{fig:doubling}). Figure~\ref{fig:sb2fit} shows one such system recovered from a real DR19 spectrum, with the single- and two-component fits and the implied component decomposition. We optimize the three parameters by a multi-start search over $q$ and a velocity grid (Section~\ref{sec:analysis-skill}), holding the primary labels at their single-star values, and retain the maximum-likelihood composite. We run the catalog through the public fitting layer of \texttt{binspec}, which is also available through our servers. We use our network in place of the \citetalias{2018MNRAS.476..528E} line network. This fitting layer refits the primary labels and allows each component to have its own $v_{\rm macro}$. We keep the other networks released with \texttt{binspec}. These place the secondary on the isochrone and give the radius and continuum shape of each component.

The detection statistic is the resulting improvement in fit,
\begin{equation}
\dchi = \chi^2_{\rm single} - \chi^2_{\rm binary}
\end{equation}
together with the improvement fraction \citep[Eq.~B1 of][]{2018MNRAS.476..528E},
\begin{equation}
\fimp = \frac{\sum_\lambda \left(|f_{\rm single}-f| - |f_{\rm bin}-f|\right)/\sigma_\lambda}{\sum_\lambda |f_{\rm single}-f_{\rm bin}|/\sigma_\lambda},
\end{equation}
where $f$ is the observed flux, $f_{\rm single}$ and $f_{\rm bin}$ the best-fit single-star and binary models, $\sigma_\lambda$ the per-pixel uncertainty, and the sums run over all good pixels. The numerator is the reduction in absolute residual the binary model buys over the single-star model, and the denominator is the total difference between the two models, so $\fimp\to1$ when the binary model's departure from the single-star model lands on the data across many pixels. A genuine SB2 improves the fit broadly, across the many line cores the secondary shifts, and so has a large $\fimp$. A fake improvement driven by a handful of outlying pixels far from any line has a small one, and is rejected by the improvement-fraction floor of the acceptance ladder. Acceptance is defined by the sliding ladder of Section~\ref{sec:acceptance}, with the rung values recalibrated for the classifier.

The same decomposition operates on the individual visit spectra as well as on the combined spectrum, which matters because the two views are sensitive to different systems. A wide or long-period binary whose components barely move over the survey baseline is line-doubled in the coadd and detected there. A close binary, by contrast, can have its components nearly velocity-aligned in one coadd yet well separated in the individual visits, where the orbital motion spreads them across the line profile. Such systems are the ones a single-epoch search misses and a multi-epoch search recovers, and they overlap the radial-velocity-variable population that dominates the close-binary catalogs of APOGEE \citep{2020ApJ...895....2P, 2021AJ....162..184K, 2020MNRAS.499.1607M, 2022MNRAS.512.2051D, 2018AJ....156...18P}. We therefore also fit the same forward model jointly to a star's individual visits, though our main catalog below is built from the combined spectra alone.

\subsection{The acceptance criterion}\label{sec:acceptance}

A star is accepted as SB2 when its improvement $\dchi$ (single minus binary) clears one of the ladder floors of table~B1 of \citetalias{2018MNRAS.476..528E} and its improvement fraction $\fimp$ exceeds the paired minimum: from $\fimp\ge 0.225$ at $\dchi=300$ down to $\fimp\ge0$ at $\dchi\ge3000$. For the catalog the ladder values are recalibrated for our classifier, scaling both coordinates of the rungs by $0.90$ (a star passes when $\dchi/0.90$ and $\fimp/0.90$ clear the published ladder). The scale is chosen at a fixed control false-positive-rate target on one half of the held-out controls and validated on the other, which sets the $8.1\%$ validated rate of Section~\ref{sec:res-bench}. We also require the scaled $\fimp$ to be at least $0.14$ at every rung. This removes detections caused by a few pixels at the detector edges. The requirement is waived only for the extreme improvements $\dchi>10^5$, where even a small $\fimp$ reflects a pervasive real mismatch.

\begin{figure}
\centering
\includegraphics[width=\columnwidth]{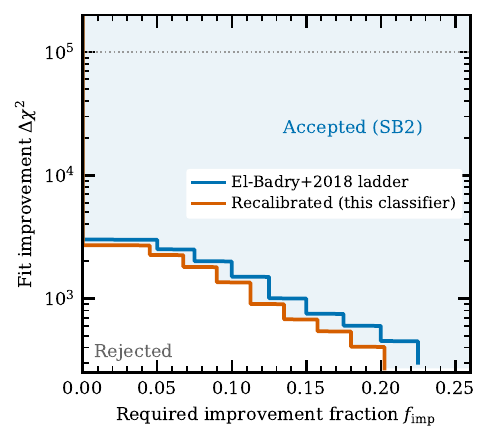}
\caption{The sliding acceptance criterion. Blue: the \citetalias{2018MNRAS.476..528E} ladder, the minimum $\fimp$ required of a star as a function of its fit improvement $\dchi$ (the largest floor it clears sets the requirement). Orange: the threshold recalibrated for our classifier (rungs scaled by $0.90$ in both coordinates), waived above $\dchi=10^5$ (gray dotted). A star is accepted as SB2 in the shaded region to the right of the recalibrated threshold.\label{fig:ladder}}
\end{figure}

For a new survey the {\it procedure}, calibrating these floors against a labeled benchmark at a fixed control false-positive rate, carries over, while the numerical values are re-derived per instrument (Section~\ref{sec:disc-instrument}).

\subsection{The operating decisions (the Skill)}\label{sec:analysis-skill}

The equations above do not by themselves reproduce the \citetalias{2018MNRAS.476..528E} catalog. What makes the method work is a set of operating decisions, and each one changes the resulting catalog in a way that the equations alone do not reveal. We record them in the Skill as an ordered list, giving for each the choice, the reason, and the failure mode it prevents, so that an agent (or a human) has the {\it why} as well as the {\it what}. All eight are written as imperative lines in the Skill, which is released with the software.

Decisions~1, 2, 5, and~8 are also enforced by servers (the shared normalization operator, the single-star model, the ladder calibration, and the isochrone-tied composite of Eq.~1), so for those the Skill line records a choice the tools also implement. In the blind-agent test, removing a server-enforced decision resets both the Skill line and the server it controls: decision~1 by dropping the self-consistent normalization and decision~8 by disabling the isochrone-tied luminosity weighting in the composite, so the agent must recover the decision from its measurable effect (Section~\ref{sec:res-blind}). The eight load-bearing decisions are the following.

\begin{enumerate}
\item {\it Self-consistent normalization.} The model and the data are normalized by the same operator, so that a continuum error enters $\chi^2_{\rm single}$ and $\chi^2_{\rm binary}$ identically and cancels in $\dchi$. Adopting the survey continuum, or normalizing the two sides separately, leaves a residual that the extra freedom of the two-component model absorbs; $\dchi$ then reads a normalization mismatch as a detection. This is the single most consequential choice.

\item {\it A five-label single-star model.} The single-star model spans $\teff$, $\logg$, $v_{\rm macro}$, and two abundance dimensions ($\feh$ and $\mgh$) rather than the first three alone, because a three-label model leaves residual line structure that the binary model can also absorb, inflating the false-positive rate near the detection threshold.

\item {\it Pixel masking and a signal-to-noise limit.} Persistently bad pixels are masked and the per-pixel signal-to-noise is capped (at 200 for APOGEE) before the fit. A handful of underestimated error bars at high signal-to-noise otherwise drive $\dchi$ upward without any broadband improvement, producing detections localized to a few pixels.

\item {\it A mass-ratio multi-start.} The fit is restarted from several trial mass ratios, because $\dchi(q)$ is multi-modal and a single start settles in the wrong minimum. In particular the $q\!\to\!1$, equal-velocity limit must not be allowed to collapse onto a single velocity-shifted single-star spectrum.

\item {\it Threshold calibration at a fixed control false-positive rate.} The sliding acceptance thresholds are calibrated against labeled real data rather than on model-into-model spectra: \citetalias{2018MNRAS.476..528E} derived their rung values from semi-empirical injections, and we re-derive the values for our classifier on real held-out controls at a fixed false-positive-rate target (Section~\ref{sec:acceptance}). The {\it procedure} transfers to a new survey; the calibrated {\it values} are re-derived per classifier and instrument (Section~\ref{sec:disc-instrument}).

\item {\it Completeness over the $(q,\teff)$ plane.} Completeness is reported as a function of mass ratio and primary temperature rather than as a single number, because the temperature floor of the single-star model raises the minimum detectable mass ratio for cool primaries.

\item {\it A positive improvement-fraction requirement.} Candidates whose $\dchi$ concentrates on a few detector-edge pixels rather than on the shared line cores that a real secondary produces are rejected by requiring $\fimp$ to clear its rung of the acceptance ladder, with the rung values calibrated for the classifier in use at a fixed control false-positive rate, and with the high-$\dchi$ waiver of Section~\ref{sec:acceptance}.

\item {\it Isochrone-tied luminosity weighting.} The flux ratio of the two components is fixed from an isochrone at the shared age and metallicity rather than fitted from the mean continuum, because a continuum-based weight biases the recovered $q$ low at the equal-mass end.
\end{enumerate}

\noindent Decisions~1 and~8 bracket the blind-agent test of Section~\ref{sec:res-blind}. The first leaves a large, label-free internal signal that an agent can act on, the second changes only a derived quantity and leaves none.

\begin{deluxetable*}{llcl}
\tablecaption{The nine MCP tool servers and what carries over to a new survey. Inputs and returns are abbreviated; every field is typed and carries units. The eight operating decisions and the seven survey-agnostic servers carry over as {\it procedures}; calibrated numbers (e.g.\ the ladder thresholds of decision~5) are re-derived per survey by the same recipe.\label{tab:servers}\label{tab:transfer}}
\tablehead{\colhead{Server} & \colhead{Representative tool: inputs $\to$ returns} & \colhead{Transfers} & \colhead{Role}}
\startdata
\texttt{data}         & \texttt{load(id)} $\to$ wl, flux, ivar, snr      & no  & read a survey spectrum \\
\texttt{single\_star} & \texttt{predict(\teff,\logg,\feh)} $\to$ flux    & no  & single-star spectral model \\
\texttt{binary\_model}& \texttt{single\_vs\_binary} $\to$ $\dchi$, $\fimp$, $q$, $v_1$, $v_2$ & yes & composite fit + statistic (Eqs.~1--2) \\
\texttt{isochrone}    & \texttt{secondary($q$,\teff)} $\to$ \teff$_2$, ratio & yes & map $q$ to the secondary, any band \\
\texttt{doppler}      & \texttt{shift(flux,$v$)} $\to$ flux              & yes & velocity shift on any grid \\
\texttt{broadening}   & \texttt{convolve(flux,$R$)} $\to$ flux           & yes & broadening at any $R$ \\
\texttt{gaia\_sql}    & \texttt{xmatch(id)} $\to$ $\varpi$, RUWE, NSS    & yes & astrometric cross-match \\
\texttt{gaia\_xp}     & \texttt{overluminous(id)} $\to$ $\Delta M_G$     & yes & color--magnitude test \\
\texttt{vision}       & \texttt{inspect(render)} $\to$ confidence, flag  & yes & read a rendered fit, veto weak \\
\hline
Skill (8 decisions)   & written operating decisions                      & yes & method-level, not instrument-specific \\
\enddata
\tablecomments{Seven of nine servers and all eight Skill decisions are reused unchanged across surveys; the data server is rewritten for the new survey's spectra and the single-star model is replaced by one tuned to the new instrument's flux calibration (Section~\ref{sec:disc-instrument}).}
\end{deluxetable*}

\begin{figure*}
\centering
\resizebox{\textwidth}{!}{
\begin{tikzpicture}[
    >={Latex[length=2.2mm]}, semithick, draw=black!60, font=\small,
    stage/.style={rounded corners=3pt, draw=black!55, thick, align=center,
        inner sep=4pt, minimum height=11mm},
    ioN/.style={stage, fill=black!10, text width=22mm, font=\footnotesize},
    agN/.style={stage, fill=oiBlue!85, text=white, text width=40mm, font=\footnotesize},
    outN/.style={stage, fill=oiGreen!80, text=white, text width=26mm, font=\footnotesize},
    skN/.style={stage, fill=black!6, text width=32mm, font=\footnotesize},
    gate/.style={diamond, aspect=1.4, draw=black!55, thick, fill=black!12,
        align=center, font=\scriptsize, inner sep=1pt},
    srvG/.style={rounded corners=2pt, draw=black!35, fill=oiGreen!78, text=white,
        font=\scriptsize, minimum height=6mm, text width=18mm, align=center, inner sep=2pt},
    srvO/.style={srvG, fill=oiOrange!82},
    flow/.style={->, semithick, draw=black!60},
    lab/.style={font=\scriptsize}
]
\node[skN]  (sk) at (0.2,2.5) {\textbf{Skill}\\ 8 operating decisions};
\node[ioN]  (in) at (-1.2,4.6) {Stellar spectrum\\ (APOGEE)};
\node[agN]  (ag) at (3.2,4.6) {\textbf{Language-model agent}\\ (Skill-primed)\\ {\scriptsize plan\,$\to$\,call\,$\to$\,observe\,$\to$\,decide}};
\node[gate] (gate) at (7.0,4.6) {Table~B1\\ $\Delta\chi^2,f_{\rm imp}$};
\node[outN] (out) at (10.4,4.6) {Single / SB2\\ $q,\Delta\chi^2,v_1,v_2$};
\draw[flow] (sk) -- (ag);
\draw[flow] (in) -- (ag);
\draw[flow] (ag) -- (gate);
\draw[flow] (gate) -- node[lab,above,text=oiGreen!55!black,pos=0.35]{Accept} (out);
\node[srvG] (s1) at (-0.45,0.7) {binary\_model};
\node[srvG] (s2) at (1.55,0.7) {isochrone};
\node[srvG] (s3) at (3.55,0.7) {doppler};
\node[srvG] (s4) at (5.55,0.7) {broadening};
\node[srvG] (s5) at (0.8,-0.4) {gaia\_sql};
\node[srvG] (s6) at (2.8,-0.4) {gaia\_xp};
\node[srvG] (s7) at (4.8,-0.4) {vision};
\node[srvO] (o1) at (7.95,0.7) {data};
\node[srvO] (o2) at (7.95,-0.4) {single\_star};
\begin{scope}[on background layer]
  \node[rounded corners=3pt, draw=oiGreen, fill=oiGreen!8, inner sep=6pt,
        fit=(s1)(s2)(s3)(s4)(s5)(s6)(s7)] (gb) {};
  \node[rounded corners=3pt, draw=oiOrange, fill=oiOrange!8, inner sep=6pt,
        fit=(o1)(o2)] (ob) {};
\end{scope}
\node[lab, text=oiGreen!55!black, anchor=north] (lg) at (gb.south) {Survey-agnostic MCP (reused)};
\node[lab, text=oiOrange!60!black, anchor=north] (lo) at (ob.south) {Replaced per survey};
\begin{scope}[on background layer]
  \node[rounded corners=5pt, draw=black!60, thick, inner sep=12pt,
        fit=(gb)(ob)(lg)(lo)(sk)] (mcp) {};
\end{scope}
\node[font=\footnotesize\bfseries, anchor=north east] at ([xshift=-10pt]mcp.north east) {Agent-callable tools};
\draw[<->, semithick, draw=black!55] (ag.south) -- node[lab,right,inner sep=2pt]{call\,/\,value\,$+$\,flag} (gb.north);
\node[ioN, text width=56mm, fill=black!6] (fb) at (7.0,6.9)
     {\footnotesize \textbf{Refine \& re-fit:}\ adjust window / re-start $q$ / mask / re-normalize};
\draw[flow, draw=oiOrange] (gate.north)
     -- node[lab, left, pos=0.55, text=oiOrange]{Borderline} (fb.south);
\draw[flow, draw=black!55] (fb.west) -- (3.2,6.9) -- (ag.north);
\begin{scope}[on background layer]
  \node[rounded corners=4pt, draw=black!45, inner sep=13pt,
        fit=(sk)(in)(ag)(gate)(out)(mcp)(fb)] (plate) {};
\end{scope}
\node[lab, anchor=south east] at ([xshift=-6pt,yshift=6pt]plate.south east) {For each star};
\node[anchor=east] at ([xshift=-0.6cm]plate.west) {\begin{tikzpicture}[y=0.72cm]
  \node[agN, minimum width=6mm, minimum height=6mm, text width=4mm, inner sep=1pt, label={[lab]right:Agent}] at (0,2) {};
  \node[srvG, minimum width=6mm, text width=4mm, label={[lab]right:Reused MCP}] at (0,1) {};
  \node[srvO, minimum width=6mm, text width=4mm, label={[lab]right:Replaced MCP}] at (0,0) {};
\end{tikzpicture}};
\end{tikzpicture}
}
\caption{The agent and its tools, as a processing flow. A Skill-primed language-model agent fits each spectrum with a single-star and a two-component model, reads the detection statistic ($\dchi$, $\fimp$) against the \citetalias{2018MNRAS.476..528E} acceptance gate, and, for borderline cases, inspects the rendered fit and \Gaia\ evidence before deciding single versus SB2. Borderline or flagged fits are refined and re-fit before the final accept-or-reject verdict. Green marks the seven servers and the operating decisions that are survey-agnostic and reused unchanged. Orange marks the two components replaced per survey, the data server and the single-star model (Section~\ref{sec:disc-instrument}).\label{fig:agentic}}
\end{figure*}
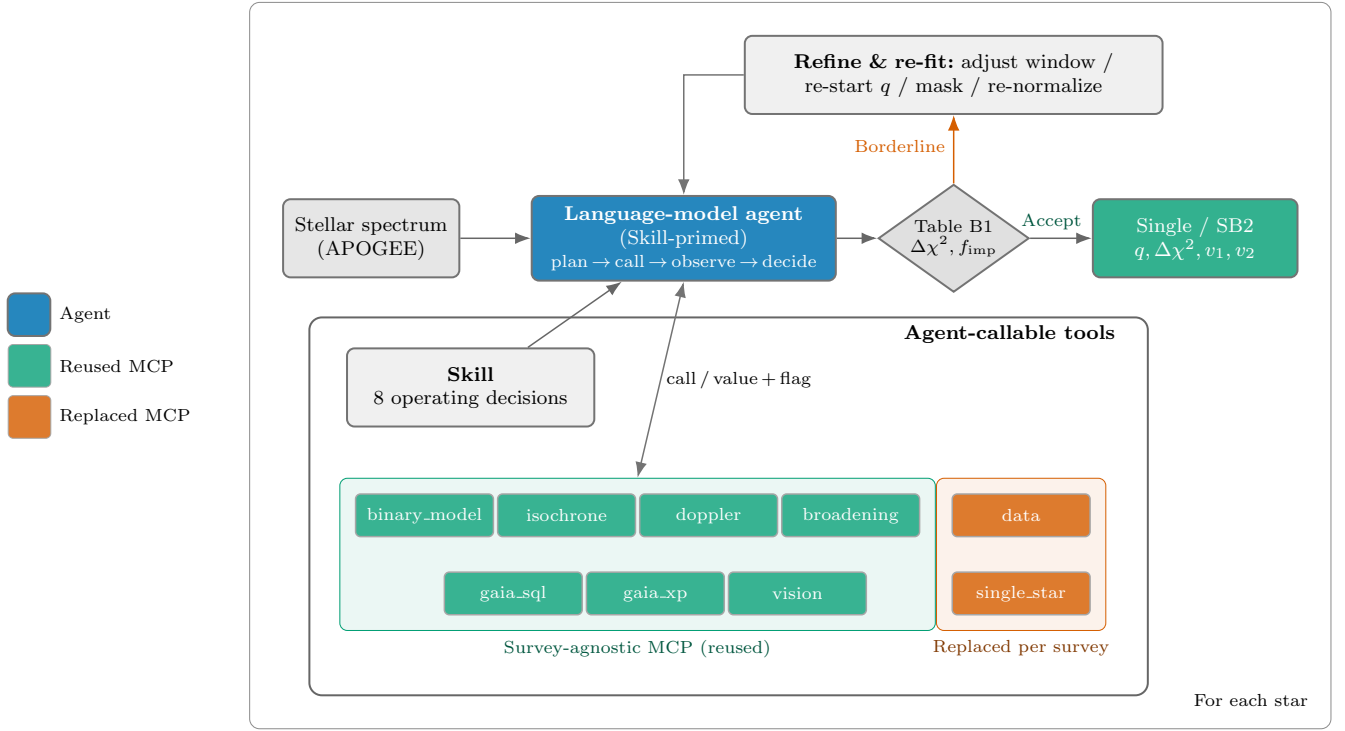

\subsection{The MCP servers and the agent}\label{sec:analysis-system}

We expose these capabilities as nine MCP servers (Table~\ref{tab:servers}; Figure~\ref{fig:agentic}). Each server declares typed tools with units on every input and output and returns a compact value rather than a file the agent parses. The science behind them lives in a deterministic module that returns identical results under any agent or under a fixed script, so reproducibility rests on the tools and the Skill rather than on the specific language model.

\begin{figure}
\centering
\resizebox{0.82\columnwidth}{!}{%
\begin{tikzpicture}[
  call/.style={draw=oiBlue, rounded corners=2pt, fill=oiBlue!7, font=\scriptsize,
    text width=34mm, align=left, inner sep=2.2pt, minimum height=6mm},
  ret/.style={draw=oiGray, rounded corners=2pt, fill=black!3, font=\scriptsize\ttfamily,
    text width=34mm, align=left, inner sep=2.2pt, minimum height=6mm},
  sflow/.style={-{Latex[length=1.8mm]}, semithick, oiGray},
  node distance=3.6mm,
]
\node[draw=oiBlue, very thick, rounded corners=3pt, fill=oiBlue!12, font=\footnotesize\bfseries,
  text width=72mm, align=center, minimum height=7mm] (top) {Agent: Is this dwarf single or SB2?};
\node[call, below=of top] (c1) {\textbf{Call} \textsf{data.load}(sdss\_id)};
\node[ret, below=of c1] (r1) {7105 px, SNR 224};
\node[call, below=of r1] (c2) {\textbf{Call} \textsf{binary\_model.single\_vs\_binary}};
\node[ret, below=of c2] (r2) {$\Delta\chi^2$=33095, $f_{\rm imp}$=0.43,\\ q=0.91, prefers\_binary=T};
\node[call, below=of r2] (c3) {\textbf{Call} \textsf{isochrone.consistent}(q, Teff)};
\node[ret, below=of c3] (r3) {consistent = True};
\node[call, below=of r3] (c4) {\textbf{Call} \textsf{gaia\_xp.overluminous} / ruwe};
\node[ret, below=of c4] (r4) {RUWE 1.20, no NSS solution};
\node[call, below=of r4] (c5) {\textbf{Call} \textsf{vision.inspect}(fit render)};
\node[ret, below=of c5] (r5) {Doubled line cores seen};
\node[draw=oiGreen, very thick, rounded corners=3pt, fill=oiGreen!12, font=\footnotesize\bfseries,
  text width=72mm, align=center, minimum height=7mm, below=of r5] (dec) {Decide: SB2 (record $q$, $\Delta\chi^2$, velocities)};
\foreach \a/\b in {top/c1,c1/r1,r1/c2,c2/r2,r2/c3,c3/r3,r3/c4,c4/r4,r4/c5,c5/r5,r5/dec}
  \draw[sflow] (\a) -- (\b);
\end{tikzpicture}
}
\caption{One agent trajectory on a flagged SB2. The agent (blue) calls the tools in order and reads back each compact return (gray), then issues the verdict (orange). The values shown are those the catalog classifier recovers for the system in Figure~\ref{fig:sb2fit}, together with its \Gaia\ DR3 RUWE.\label{fig:trace}}
\end{figure}
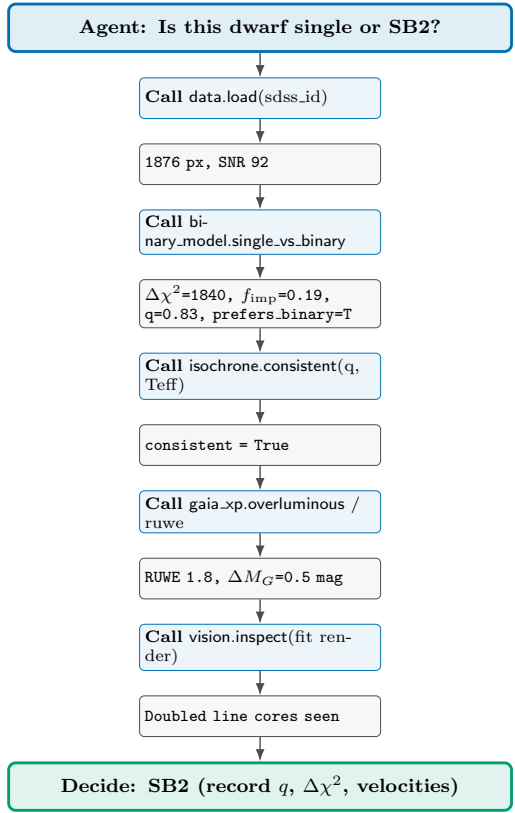

The agent is a language model primed by the Skill and connected to the servers through a plan--call--observe--decide loop, implemented as a small state graph in LangGraph. The graph has one node per stage. A planning node reads the operating decisions and chooses which tools to call and in what order. A tool-calling node invokes a server through its typed MCP interface and writes the compact return into the shared state. An observation node updates the running hypothesis (single versus binary, and the current best $q$ and velocities). A decision node either requests another tool call or emits the verdict. The design follows the reason-and-act pattern of \citet{2022arXiv221003629Y} and the learned tool-use of \citet{2023arXiv230204761S}, specialized to a fixed scientific task and constrained by the Skill so that the model's latitude is in {\it which} evidence to gather and {\it when} to stop, not in the numerical results, which come from the deterministic servers.

A vision node reads the rendered single-star and binary fits together with the \Gaia\ evidence and flags detections that do not survive inspection, catching near-equal-mass systems and blends that pass the numeric statistics. This is the one node that calls a vision-language model rather than the deterministic core, and it serves the worked examples and the benchmark audit. The full-sample catalog run is deterministic and does not use it. Adding a capability is one server plus one node in the loop, so the system grows by units rather than by rewriting a pipeline. Figure~\ref{fig:trace} shows one agent trajectory, from the first data call to the final SB2 verdict.

\section{Results}\label{sec:results}

\subsection{Benchmark recovery}\label{sec:res-bench}

Before applying the classifier to the full sample, we validate it on the benchmark of Section~\ref{sec:data-valid}, the $2{,}344$ \citetalias{2018MNRAS.476..528E}-flagged SB2 and $7{,}866$ matched single-star controls. On this labeled set the two decision statistics can be read against known truth, and recovery can be measured at a controlled false-positive rate, before the classifier is trusted on the unlabeled full sample.

\begin{figure}
\centering
\includegraphics[width=\columnwidth]{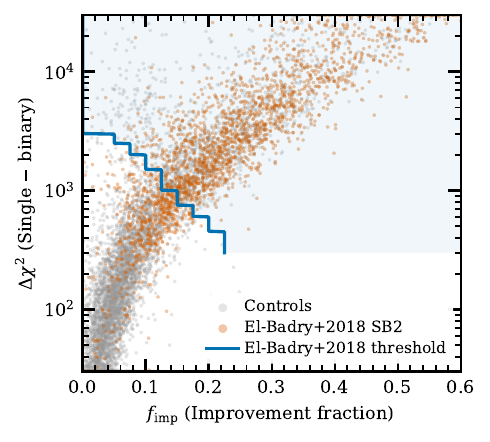} \caption{The detection statistic on the benchmark. Each point is one star in the plane of the fit improvement $\dchi$ (single minus binary) and the improvement fraction $\fimp$. Control stars (gray) stay at low values, while the benchmark SB2 (orange) climb into the acceptance region beyond the \citetalias{2018MNRAS.476..528E} threshold (blue). These are the two numbers the agent reads back from the binary-model server.\label{fig:detstat}}
\end{figure}

Figure~\ref{fig:detstat} shows the two-number decision on the benchmark. Control stars concentrate at low improvement, while the benchmark SB2 extend along the sliding \citetalias{2018MNRAS.476..528E} ladder into the acceptance region.

\begin{figure}
\centering
\includegraphics[width=\columnwidth]{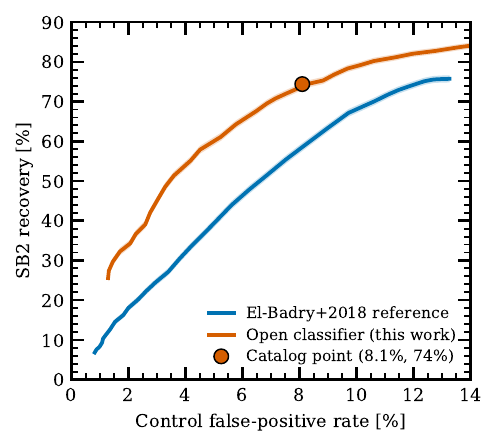}
\caption{SB2 recovery versus control false-positive rate on the benchmark ($2{,}344$ SB2 and, for the false-positive rate, the $3{,}590$ held-out controls the classifier never trained on). Our open real-data classifier (orange) is shown alongside the \citetalias{2018MNRAS.476..528E} network run as a best-case reference (blue). The adopted catalog operating point is marked ($8.1\%$ validated false-positive rate, $74.4\%$ recovery); at that false-positive rate the best-case reference recovers about $55\%$, and at its own loosest setting ($13.3\%$) it reaches $75.6\%$, which the open classifier exceeds at $82.6\%$ recovery and $12.3\%$. Bands are $1\sigma$ binomial.}\label{fig:roc}
\end{figure}

Figure~\ref{fig:roc} shows recovery versus the control false-positive rate on the 2{,}344-SB2 benchmark. Both the training of the classifier and the calibration of its acceptance thresholds are validated on held-out data. The network is trained on half of the eligible dwarfs and its false-positive rate is measured on controls it never saw, and the recalibrated threshold values are set on one half of those held-out controls and verified on the other.

At the operating point we adopt for the catalog, with the rung values recalibrated for this classifier at a fixed control false-positive-rate target (decision~5), the open real-data classifier recovers $74.4\%$ of the benchmark SB2 at an $8.1\%$ validated false-positive rate. At its loosest setting it reaches $82.6\%$ at $12.3\%$. Run as a best-case reference on the same benchmark, the \citetalias{2018MNRAS.476..528E} network reaches $75.6\%$ at its loosest $13.3\%$ point and only about $55\%$ at the catalog's $8.1\%$ false-positive rate, so the open classifier exceeds the reference at every matched false-positive rate.

Neither classifier recovers every \citetalias{2018MNRAS.476..528E}-flagged SB2, and not even \citetalias{2018MNRAS.476..528E}'s own network run as a best-case reference does. The reason is that the benchmark positives were flagged on the DR13 spectra, whereas we test on the independent DR19 re-reduction of the same stars. The two releases combine a different set of visits and pass through a different pipeline, so a star's DR19 combined spectrum can carry a smaller velocity split, lower signal-to-noise, or a near-conjunction coaddition that no longer shows resolved line doubling. Recovery here is therefore measured relative to the DR13 flags, not as an absolute completeness, and the shortfall is a property of the changed data rather than of either classifier.

\subsection{The APOGEE DR19 SB2 catalog}\label{sec:res-apogee}

\begin{figure*}
\centering
\includegraphics[width=\textwidth]{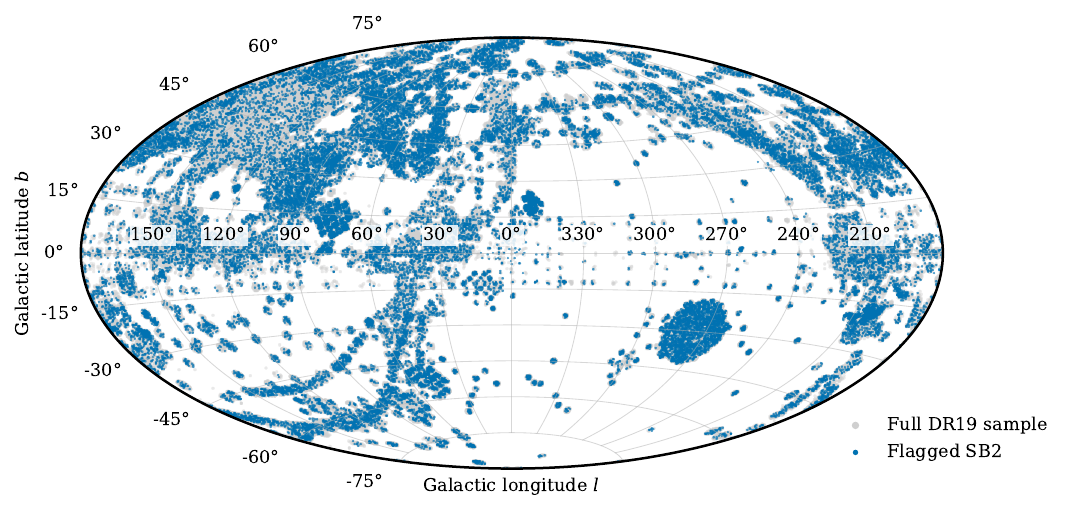}
\caption{Galactic distribution (Aitoff projection, longitude increasing to the left) of the flagged SB2. Gray: the full DR19 dwarf sample with a coordinate match in the DR19 astra ASPCAP summary; blue: the 41{,}290 flagged SB2 with a match (of 41{,}466). The flagged binaries trace the survey's disk, bulge, and halo fields, with no obvious spatial selection of their own.}\label{fig:sky}
\end{figure*}

\begin{figure}
\centering
\includegraphics[width=\columnwidth]{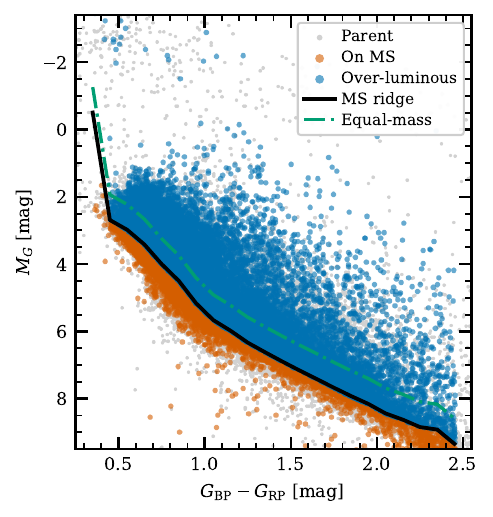}
\caption{\Gaia\ color--magnitude check of the flagged SB2, in observed (not de-reddened) $M_G$ and $G{\rm BP}-G{\rm RP}$. Gray: the full dwarf sample, whose running median is the main-sequence ridge (black solid). Blue: flagged SB2 above the ridge ($77\%$ of those with a reliable parallax), as expected for unresolved binaries. Orange: flagged SB2 on or below the ridge. The green dash-dot line is the equal-luminosity locus ($0.752$~mag above the ridge), where an equal-mass pair would sit.}\label{fig:cmd}
\end{figure}

Run over the 238{,}205 dwarfs under a fixed driver script, the classifier flags 41{,}466 SB2 (17.4\% of the full sample) at the recalibrated acceptance (Section~\ref{sec:res-bench}), which we take as the DR19 SB2 catalog, reporting the detections directly as \citetalias{2018MNRAS.476..528E} did. As set out below, the control false-positive rate implies that close to $40\%$ of these are falsely flagged singles, so the sample is a candidate list rather than a pure catalog. Table~\ref{tab:catalog} lists the released columns.
\begin{table}
\centering
\footnotesize
\setlength{\tabcolsep}{1.5pt}
\caption{Released columns of the DR19 SB2 catalog and its supplements. The rows starting with \texttt{n\_visits} come from the multi-epoch supplement. The \texttt{teff2} values come from the component-temperature table. The \Gaia\ quantities are linked through \texttt{gaia\_dr3\_source\_id}. Full descriptions and example rows accompany the data release.\label{tab:catalog}}
\begin{tabular}{lll}
\hline
Column & Unit & Description \\
\hline
\texttt{sdss\_id} & --- & SDSS-V identifier \\
\texttt{gaia\_dr3\_source\_id} & --- & \Gaia\ DR3 identifier \\
\texttt{ra}, \texttt{dec} & deg & J2000 coordinates \\
\texttt{teff\_single} & K & $\teff$ from the single-star fit \\
\texttt{logg\_single} & dex & $\logg$ from the single-star fit \\
\texttt{feh\_single} & dex & $\feh$ from the single-star fit \\
\texttt{best\_q} & --- & $q$ from the composite model \\
\texttt{teff2} & K & secondary $\teff$ (isochrone tie) \\
\texttt{best\_rv1}, \texttt{best\_rv2} & \kms & component velocities (coadd) \\
\texttt{delta\_chi2} & --- & $\dchi$ (single $-$ binary) \\
\texttt{f\_imp} & --- & improvement fraction \\
\texttt{sb2} & flag & accepted as SB2 (recalibrated gate) \\
\texttt{teff\_floor} & flag & primary at the $4200$~K model floor \\
\texttt{feh\_metalpoor} & flag & pipeline $\feh<-0.9$ \\
\texttt{coadd\_only} & flag & split not recovered in the visits \\
\texttt{vmacro\_single} & \kms & $v_{\rm macro}$ from our single-star fit \\
\texttt{vmacro\_1}, \texttt{vmacro\_2} & \kms & $v_{\rm macro}$ for each component \\
\texttt{n\_visits} & --- & number of APOGEE visits fit \\
\texttt{prefers\_binary} & flag & two stars preferred by the visit fit \\
\texttt{v1\_range} & \kms & max.\ primary velocity change \\
\texttt{q\_dyn}, \texttt{q\_wilson} & --- & dynamical mass ratios \\
\texttt{sb1} & flag & velocity variable outside catalog \\
\texttt{v1\_span\_gt20} & flag & $\ge3$ nights spanning $>20~\kms$ \\
\hline
\end{tabular}
\end{table}

The flagged systems span the full APOGEE disk, bulge, and halo footprint, tracing the full dwarf sample with no spatial selection beyond that of the survey itself (Figure~\ref{fig:sky}). This detection rate is comparable to the 13.1\% ($2{,}645$ of $20{,}142$) of \citetalias{2018MNRAS.476..528E}, reached at the benchmark-controlled false-positive rate of Section~\ref{sec:res-bench} (Figure~\ref{fig:roc}). The median composite-model mass ratio is $q=0.91$, or $q=0.80$ over the $0.2$--$0.95$ interval where the decomposition is most sensitive, close to the $0.83$ of \citetalias{2018MNRAS.476..528E}, where $q$ is also derived from this kind of composite model.

As an independent check, following \citetalias{2018MNRAS.476..528E}, we place the flagged systems in the \Gaia\ color--magnitude diagram (Figure~\ref{fig:cmd}). Of the $36{,}300$ with a parallax of signal-to-noise $\varpi/\sigma_\varpi>5$, $77\%$ lie above the single-star main sequence, the over-luminosity expected of an unresolved binary. We cross-matched a random sample of $6{,}000$ dwarfs from the search to \Gaia\ DR3. In this sample, the flagged stars have higher astrometric noise ($49\%$ with RUWE$>1.4$, against $19\%$ of the unflagged stars) and more non-single-star solutions ($16\%$ versus $5\%$). APOGEE DR17 also observed $29{,}233$ of the flagged systems. Of these, $11\%$ are listed as SB2 in \citet{2021AJ....162..184K}. This fraction is a lower bound set by that catalog's velocity-separation selection. It increases with the coadd separation, from $3\%$ below $10~\kms$ to $36\%$ above $30~\kms$. These independent tracers corroborate the catalog.

A direct purity fraction awaits dedicated spectroscopic follow-up. The identifiable contaminants are minor, and both are quoted as fractions of the $41{,}466$ flagged systems, following the \citetalias{2018MNRAS.476..528E} estimates. Chance alignments within a fiber contribute $\lesssim$1\% of the flagged sample and hierarchical triples mis-fit as two components contribute $\sim$3\% of it. One example is the known triple \Gaia\ DR3 151001777293928576, which our two-component model fits poorly but still flags (Section~\ref{sec:res-multiepoch}, Figure~\ref{fig:refstar}). Both contaminants land in the catalog at these enhanced rates, rather than at their much lower parent-sample rates, because both produce the near-equal-luminosity composites the classifier selects for. Together they are an order of magnitude below the spectroscopic false-positive term that follows. Read at face value, the $8.1\%$ held-out control false-positive rate applies to the single-star population, which the $\sim$$197{,}000$ unflagged dwarfs approximate, and implies of order $16{,}000$ falsely flagged singles if the control rate translates directly to the full sample. Those land inside the catalog, where they would make up close to $40\%$ of the $41{,}466$ flagged systems. The $77\%$ over-luminous fraction (against about half for ridge-like stars by construction of the ridge) and the $46\%$ multi-epoch confirmation (Section~\ref{sec:res-multiepoch}), compared with $94\%$ for SB2 also identified in published surveys, are both consistent with contamination of that order. An external check that does not condition on the spectra points the same way. On a random subsample of the searched dwarfs cross-matched to \Gaia, the flag rate among astrometrically quiet stars, RUWE below $1.1$ and no non-single-star solution, is $9.6\pm0.5\%$. Real short-period binaries produce little astrometric wobble and survive that cut, so this rate is an upper bound on the false-positive rate, and it brackets the $8.1\%$ control value from the independent side.

One artifact of the coadd deserves comment. APOGEE shifts each visit to a common frame with a single-star cross-correlation velocity, which for a real SB2 is measured on a blended profile, so the component velocities we report from the coadd are splittings about the pipeline rest frame rather than absolute velocities. For $10{,}000$ unflagged dwarfs, the single-star fits to the coadds give a median velocity of $+0.03~\kms$ relative to the rest frame. The robust scatter is $0.09~\kms$, compared with $0.40~\kms$ for the flagged SB2. The multi-epoch refit never touches the coadd, and we treat its per-visit velocities as the more reliable of the two. Some coadds look composite even though their visits do not. We use the flag \texttt{coadd\_only} for a catalog SB2 with at least two fitted visits and a coadd separation above $15~\kms$. The pipeline velocity must also be steady, with a scatter below $1~\kms$, and the visit fit must prefer one star. A real pair with this separation shows two sets of lines in every visit. For published SB2, the visit fit prefers two stars in $92$ to $95\%$ of cases at every coadd separation. The flag marks $2{,}239$ systems, or $8.5\%$ of the catalog systems with multiple visits. It marks $1.4\%$ of the SB2 also identified in published surveys. Two of the stars marked by the flag are \texttt{sdss\_id} 61585132 and 76096409, and the visit fit prefers one star for both. Their DR19 pipeline velocities agree within $0.5~\kms$ across visits. For 76096409, this holds even though the barycentric correction changes by $12~\kms$, which shows that the visits were combined in the correct frame. Both stars are metal-poor, as are $942$ of the flagged systems. Among metal-poor stars, the component away from the pipeline velocity has a positive velocity in $90\%$ of cases. For the rest of the sample, the fraction is $52\%$. This points to the single-star model as the source of these composite fits, rather than the coadd. Users of the DR19 \texttt{the\_payne\_apogee\_star} parameters can use this flag to remove these stars from the SB2 list and keep them as single stars.

The control sample itself retains binaries, which the \Gaia\ astrometry shows. Our classifier flags the held-out controls with a non-single-star solution at $14.4\%$ and those with ${\rm RUWE}>1.4$ at $13.0\%$, against $7.7\%$ for the astrometrically quiet ones, so the $8.1\%$ control rate is an upper bound on the true false-positive rate. We release the benchmark with \Gaia\ identifiers and these flags so the cut can be repeated.

Published SB2 catalogs bound the same effect from outside. We cross-match the controls against \citet{2021AJ....162..184K}, \citet{2022MNRAS.517..356K} and \citet{2024MNRAS.527..521K} and find $125$ stars ($1.59\%$), of which our preliminary binary pass had already removed $62$. Of the rest, $28$ entered the training set of the catalog network and $35$ fall in the held-out controls, and dropping those moves the held-out control false-positive rate from $8.1\%$ to about $7.8\%$. We release the cross-match with the benchmark table.

Metal-poor primaries deserve separate caution. Stars with pipeline $\feh<-0.9$ make up $3.3\%$ of the searched sample but $10.4\%$ of the flagged one, and their control false-positive rate is $38\%$ against $14\%$ for the rest, since we train the single-star model on a predominantly near-solar sample and the second component absorbs the mismatch. The multi-epoch refit fits the same single-star model to every visit and does not remove this failure, so we mark these systems in the release rather than treat them as confirmed.

Users who need a higher-purity sample should select on the multi-epoch confirmation flags of the released supplement, or at least exclude systems marked as \texttt{coadd\_only}. The counts in this paper refer to the full flagged sample, reported at its stated false-positive rate as in \citetalias{2018MNRAS.476..528E}.

\subsection{Properties of the catalog}\label{sec:res-catalog}

\begin{figure*}
\centering
\includegraphics[width=\textwidth]{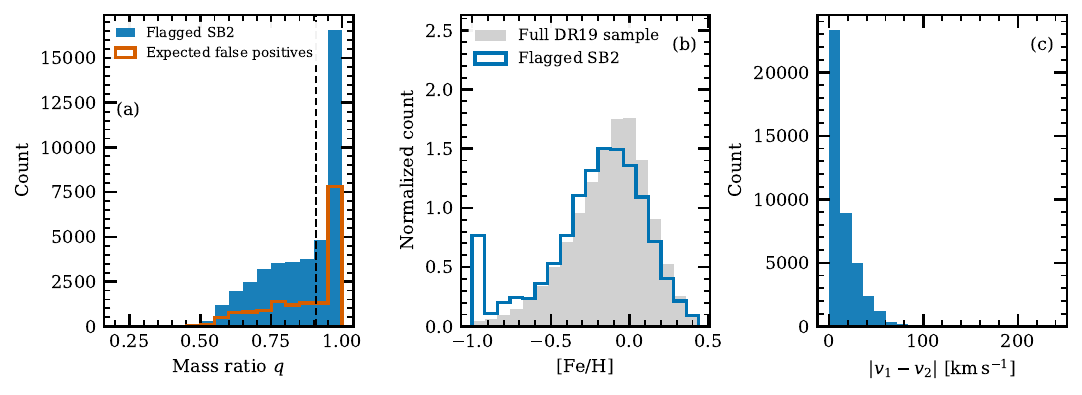}
\caption{Properties of the 41{,}466 flagged DR19 SB2. (a) recovered mass ratio $q$, shown for $q\ge0.2$ (median dashed), with the $q$ distribution of the falsely flagged benchmark controls, which we scale to the expected $\sim$$16{,}000$ false positives and overlay in orange. (b) primary metallicity of the flagged SB2 (blue) against the full DR19 dwarf sample (gray), area-normalized; the raw SB2 fraction versus $\feh$ is selection-dominated and not shown. The pile-up at $\feh=-1$ is a reporting floor rather than a metal-poor excess, since we train the isochrone-tie network only down to $\feh=-0.99$ and set the lower fitting limit to $\feh=-1$, and more metal-poor primaries, with pipeline values running to $-1.5$ and a median of $-1.36$, are clipped there. (c) the implied velocity separation $|v_1-v_2|$ of the two components in the combined spectrum. All quantities are as recovered by the classifier, not completeness-corrected.}\label{fig:catalog}
\end{figure*}

Figure~\ref{fig:catalog} summarizes the catalog. The recovered mass-ratio distribution (panel a) increases toward equal masses, with a peak at the near-equal-mass twins. Part of that peak is expected, since an excess of near-equal-mass systems is established for close binaries \citep{2017ApJS..230...15M, 2019MNRAS.489.5822E}. Part of it is contamination. The falsely flagged benchmark controls concentrate at high mass ratio, with a median $q$ of $0.94$ and half above $0.95$, because a near-equal composite at small velocity separation is the direction in which the two-component model degenerates toward a single star. Panel a overlays their distribution scaled to the expected number of false positives, which accounts for close to half of the $q>0.95$ bin. The height of the twin peak should be read with that overlay subtracted in mind. The shape is nonetheless not the intrinsic mass-ratio function, being set also by the classifier's $q$-dependent completeness and, toward $q\to1$, by the small velocity split of zero-offset twins. The fitted mass ratio is a parameter of the composite model. It comes from mapping the flux ratio of the two components to mass using a single 4~Gyr isochrone, so it should not be treated as a measured mass ratio. We refitted every released SB2 from new starting points. This changes $q$ by more than $0.1$ for $6.4\%$ of the systems. In the same refit, $2{,}647$ systems ($6.4\%$) fall below the acceptance gate. We therefore interpret the distribution of fitted mass ratios as a property of the classifier. Where the visits allow it, we compare these values with the dynamical mass ratios (Section~\ref{sec:res-multiepoch}), which are the measurements published by \citet{2024MNRAS.527..521K} for LAMOST-MRS. \citet{2022MNRAS.510.1515K} and \citet{2022MNRAS.517..356K} also found disagreement between the spectroscopic and dynamical mass ratios.

We keep the near-equal-mass twins in the catalog. At $q>0.95$ the flagged systems have a median velocity separation of $11~\kms$ and share the fit statistics of the rest of the sample, so they are line-doubled rather than degeneracies. What stays out of reach is the zero-velocity-offset twin. An equal-temperature pair with no velocity split in the combined spectrum is indistinguishable from a single star there. A close pair whose individual visits show velocity variation is instead recoverable from its per-visit velocities. Only a twin aligned in velocity at every epoch stays unrecoverable, which is why the median-mass-ratio comparison with \citetalias{2018MNRAS.476..528E} (Section~\ref{sec:res-apogee}) is made over the $0.2$--$0.95$ range.

In metallicity (panel b) the flagged SB2 track the full DR19 dwarf sample closely, with median pipeline $\feh=-0.15$ against $-0.11$ for the full sample. We do not interpret this as evidence for or against the metallicity dependence of the close-binary fraction reported from radial-velocity variability \citep{2018ApJ...854..147B, 2020MNRAS.499.1607M}, because the observed SB2 fraction as a function of $\feh$ is strongly shaped by the temperature floor and by the $\feh$-dependent completeness of the classifier. A completeness-corrected multiplicity--metallicity measurement requires the two-dimensional selection function we defer to future work.

The velocity separation of the two components in the combined spectrum (panel c) has a median $|v_1-v_2|=9.5~\kms$ and a tail to $\sim$36~$\kms$ at the $90$th percentile. This is the separation at the single epoch of the combined spectrum, not an orbital amplitude. Systems caught near conjunction, where the components are nearly aligned in velocity, are recovered through the line-doubling that the composite model captures even at small separation, and are exactly the systems that a single-epoch cross-correlation search would miss \citep{1994ApJ...420..806Z}. Recovering the orbits themselves requires multi-epoch velocities \citep{2017ApJ...837...20P, 2004A&A...424..727P}.

\begin{figure}
\centering
\includegraphics[width=\columnwidth]{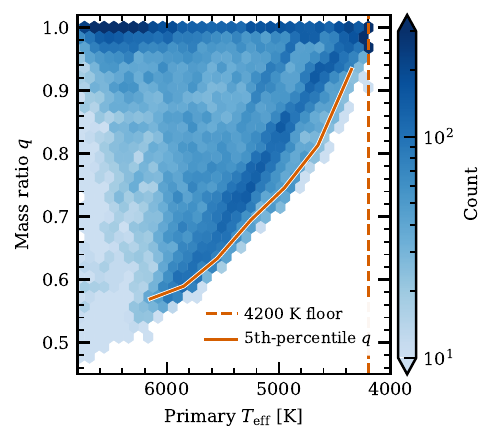}
\caption{Density of the 41{,}466 flagged SB2 in primary temperature and recovered mass ratio ($T_{\rm eff}$ increasing to the left; logarithmic color scale). The single-star model's 4200~K floor bounds the recovered mass ratio toward cool primaries (dashed). The fifth-percentile $q$ in bins of $\teff$ (solid) traces the detectability floor: cooler primaries admit only higher-$q$ companions, because the model's declining fidelity toward the floor raises the mass ratio at which a secondary still yields a decisive $\dchi$.}\label{fig:qteff}
\end{figure}

The interplay of these quantities is set by the classifier's sensitivity, which Figure~\ref{fig:qteff} makes explicit in the plane of primary temperature and mass ratio. Warm primaries populate the full range of recovered $q$, but as the primary cools the minimum recovered $q$ rises. The fifth-percentile $q$ climbs from $\approx0.6$ near 6000~K toward the reliability edge below 4700~K. This is not an astrophysical deficit of low-$q$ companions but the detectability floor of the single-star model, which loses fidelity toward its 4200~K limit so that only the brightest, most nearly equal secondaries still produce a decisive $\dchi$. The same trend appears in the injection-recovery completeness set (Section~\ref{sec:data-valid}), and it is the reason we report completeness over the $(q,\teff)$ plane rather than as a single number.

\begin{figure}
\centering
\includegraphics[width=\columnwidth]{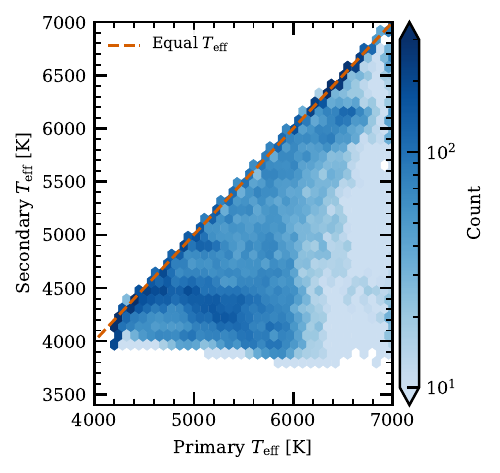}
\caption{Component temperatures of the 41{,}466 flagged DR19 SB2: the primary $\teff$ from the single-star fit against the secondary $\teff$ implied by the recovered mass ratio through the isochrone tie (logarithmic color scale). The secondaries lie on a cooler main sequence below the equal-temperature line (dashed), and the lower envelope lifts toward hotter primaries where a cool secondary contributes too little $H$-band light to be detected.}\label{fig:compteff}
\end{figure}

The mass ratios translate, through the same isochrone tie the classifier uses, into component temperatures. Figure~\ref{fig:compteff} shows the primary temperature against the secondary temperature implied by the recovered $q$. The secondaries populate a cooler main sequence below the equal-temperature line, with a median near $4{,}590$~K against $5{,}505$~K for the primaries, so the catalog is predominantly pairs of Sun-like and cooler K-type dwarfs. The lower envelope lifts toward hotter primaries for the same reason the $q$ floor does, since a cool secondary on a warm primary contributes too little $H$-band light to register.

\begin{figure*}
\centering
\includegraphics[width=\textwidth]{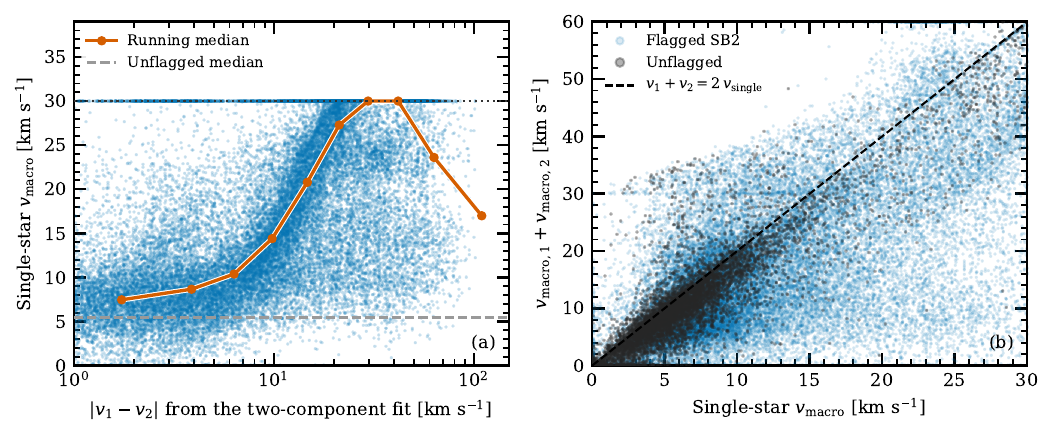}
\caption{Broadening parameters from our fits to the flagged SB2 and $10{,}000$ searched stars that were not flagged. Panel (a) compares the single-star $v_{\rm macro}$ with the component separation from the two-component fit. The two increase together up to a separation of about $30~\kms$, with a Spearman correlation of $0.70$. At this separation, the median broadening reaches the network limit of $30~\kms$, shown by the dotted line. The two no longer track each other once the line cores are resolved. The dashed line shows the median for the unflagged stars. The fit allows values up to $45~\kms$, but the model spectrum does not change above $30~\kms$. Panel (b) compares the sum of the component broadening parameters with the single-star value. This is the equivalent of the $v\sin i$ test in \citet{2022MNRAS.517..356K}. Unflagged stars are shown in black. A single star fitted with two identical components would fall on the dashed line. We exclude stars at the $30~\kms$ limit. Component values above $30~\kms$ are plotted at $30$.\label{fig:vmacro}}
\end{figure*}

The broadening the single-star fit reports can help identify a binary, since a fit forced onto an unresolved pair spends its broadening freedom on the velocity split. In these one-dimensional models, $v_{\rm macro}$ is an empirical broadening parameter. At this resolution, it is difficult to separate macroturbulence from $v\sin i$. For slowly rotating dwarfs, $v_{\rm macro}$ therefore represents the total line broadening. This is similar to $v_{\rm broad}$ in \Gaia\ DR3 \citep{2023A&A...674A...8F}, which follows $v\sin i$ \citep{2022MNRAS.517..356K}. It does not describe a property of the binary. We keep the name $v_{\rm macro}$ because it is the label used in the \citetalias{2018MNRAS.476..528E} model that we run. Figure~\ref{fig:vmacro}a shows the median $v_{\rm macro}$ climbing from the control value of $5.5~\kms$ to the $30~\kms$ network ceiling near $|v_1-v_2|\simeq30~\kms$ and falling once the cores resolve, the behavior \citet{2022MNRAS.517..356K} and \citet{2024MNRAS.527..521K} report for LAMOST-MRS. We apply their test in Figure~\ref{fig:vmacro}b. The flagged SB2 fall below $v_{\rm macro,1}+v_{\rm macro,2}=2\,v_{\rm macro}$ in $75\%$ of cases, compared with $71\%$ for the unflagged stars. The single-star $v_{\rm macro}$ reaches its limit at $30~\kms$. The composite model can also fit many single stars with two narrower components. So this test gives only a weak separation between the two samples. A high $v_{\rm macro}$ marks a candidate even where the split is too small to resolve, so we release it with the catalog, using the value from our single-star fit instead of the DR19 pipeline value, along with the values for both components, and note that $17.7\%$ of the flagged systems reach the $30~\kms$ limit, where the value is a lower bound. The network was trained on spectra with $v_{\rm macro}\le30~\kms$, and its output does not change above this value. The fit bound is $45~\kms$, so the released values between $30$ and $45~\kms$ only show that the broadening reaches the $30~\kms$ limit.

\subsection{The multi-epoch analysis}\label{sec:res-multiepoch}

The combined-spectrum search detects SB2 from a single spectrum. Where a star has more than one APOGEE visit, we fit the same forward model jointly to the individual epochs, sharing one set of stellar labels across all visits and tying the two components' velocities at each epoch by momentum conservation about the shared systemic velocity \citep{2018MNRAS.476..528E}. This tests the combined-spectrum detections against the individual visits. We use one velocity for all visits in the single-star model. A single-lined star with a changing velocity can therefore also be better fitted by the two-component model. Because the joint fit sums the improvement $\dchi$ over all $N_{\rm epoch}$ visits, we scale the acceptance threshold in proportion, requiring $\dchi>300\,N_{\rm epoch}$ (the single-epoch floor times the number of visits), so that stars with many visits are not flagged from accumulated noise alone. The joint fit is run on dwarfs flagged by combined-spectrum line doubling or by visit-velocity scatter. The DR19 visit spectra are already in the rest frame of the star, and the velocity removed from each spectrum is stored. We checked this by cross-correlating each visit with its own coadd. For $99.4\%$ of $9{,}808$ single-star visits, this confirms that the spectra are in that frame. We use the stored velocity to shift each visit back, so that the fitted velocities are barycentric. Visits without a stored velocity, or with one above $400~\kms$, are excluded. We also checked $13{,}000$ candidates that the fit identifies as single stars and whose pipeline velocities are stable. Their fitted velocities agree with the median pipeline velocities, with no offset and a scatter of $0.09~\kms$. For confirmed systems with at least three visits, the changes in fitted primary velocity follow the changes in pipeline velocity. The Spearman correlation is $0.87$, rising to $0.93$ for $q<0.7$, where the pipeline follows the brighter star. We run the joint fit from several starting points, including the component velocities from each visit's separate two-component fit.

We ran this multi-epoch fit on every catalog SB2 with visit data together with the dwarfs of visit-to-visit velocity scatter $\sigma_v>1~\kms$ (from the survey per-visit radial velocities), $50{,}187$ unique candidates in all, and refit the $26{,}204$ catalog SB2 with at least two usable epochs. For each system, we use up to eight visits with the highest signal-to-noise. The released supplement gives the median, scatter, and largest change in the DR19 pipeline velocities for every candidate. The confirmation statistics below use the $26{,}204$ of these with two or more epochs, the rest being catalog SB2 with fewer than two usable visits, along with the $\sigma_v$-selected dwarfs outside the catalog that we did not refit. Across the sample the median dynamical mass ratio, from the per-visit velocity amplitudes, is $0.84$ for the confirmed systems with velocity changes, compared with the $0.91$ of the combined-spectrum catalog.

\begin{figure}
\centering
\includegraphics[width=\columnwidth]{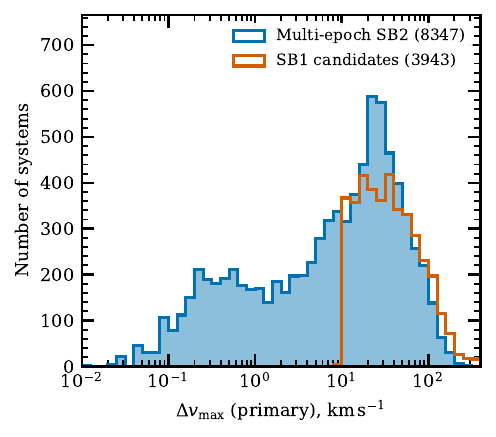}
\caption{Maximum primary radial-velocity change $\Delta v_{\rm max}$ across visits in the multi-epoch analysis, restricted to systems with three or more epochs, for which $\Delta v_{\rm max}$ is a tracer of orbital motion (two-epoch systems often sample similar orbital phases and are excluded here). The primary velocity is measured at every epoch. Blue: the $8{,}347$ catalog SB2 whose visit fits prefer two stars, whose distribution extends above $100~\kms$. The $27$ systems with changes below $0.01~\kms$ are not shown. Orange: the $3{,}943$ velocity variables (SB1 candidates), stars outside the catalog with DR19 pipeline velocity changes of $10$ to $480~\kms$, after excluding visit velocities with absolute values above $400~\kms$, the class the combined-spectrum catalog cannot flag.}\label{fig:dvmax}
\end{figure}

\begin{figure*}
\centering
\includegraphics[width=0.92\textwidth]{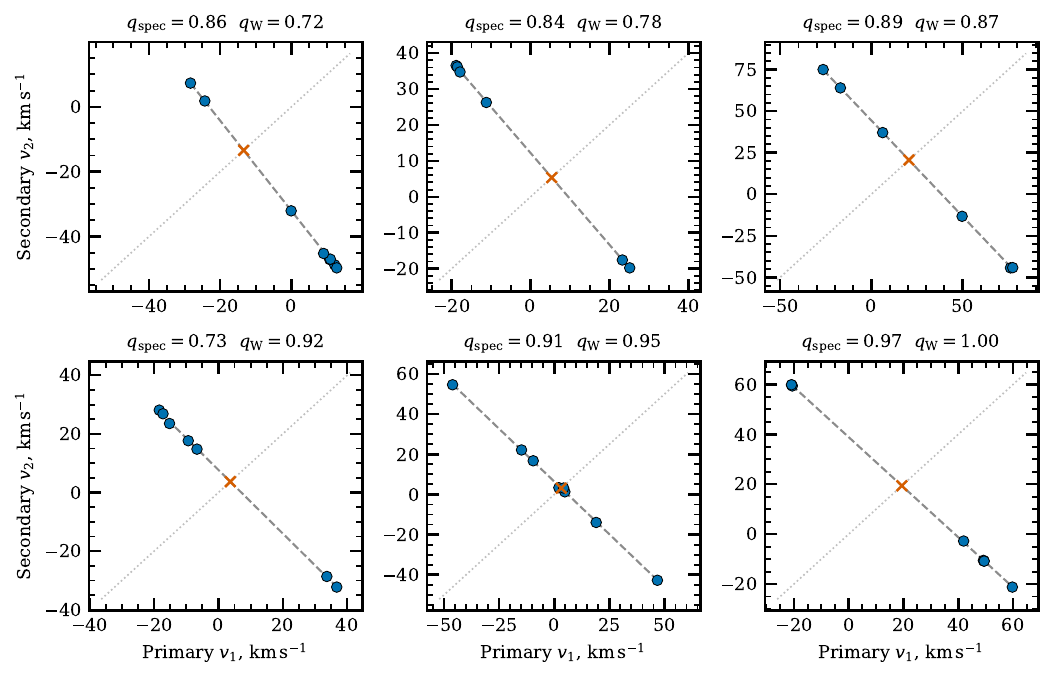}
\caption{Wilson plots for six confirmed multi-epoch SB2. Each point represents one visit. We use the component velocities from that visit's separate two-component fit, without a momentum tie. The stars move in anti-phase around the barycenter. The slope of $v_2$ against $v_1$ is $-m_1/m_2$, from which we calculate the Wilson mass ratio $q_{\rm W}$ using the dashed fit. Each panel gives $q_{\rm W}$ and the catalog's combined-spectrum \texttt{best\_q}, shown under the label qspec. This label does not refer to the supplement's \texttt{q\_spec} column. The dotted line shows $v_2=v_1$. It intersects the fit at the systemic velocity, marked by a cross. We selected these six systems from the best-sampled confirmed SB2 to cover the range of $q_{\rm W}$.}\label{fig:rvcurves}
\end{figure*}

First, the multi-epoch fit tests the combined-spectrum SB2 without using the coadd. The joint fit prefers two components for $45.5\%$ of the $26{,}204$ catalog SB2 with at least two epochs. We refer to these as confirmed systems. Those that also pass the motion cut described below are called velocity-confirmed. The confirmation rate does not depend on the number of epochs. There are $2{,}496$ systems also identified as double-lined by \citet{2021AJ....162..184K}, \citet{2022MNRAS.517..356K}, or \citet{2024MNRAS.527..521K}. The joint fit confirms $94.4\%$ of these, compared with $40.4\%$ of the remaining systems. The rate is lower for pairs with small coadd separations. It is $34.4\%$ below $15~\kms$ and $68.3\%$ between $15$ and $30~\kms$. Twins also have a lower confirmation rate, at $39.2\%$ compared with $50.2\%$ for non-twins. A pair of visits at similar orbital phase cannot show a velocity change, and a close pair may not pass the test when the visits have lower signal-to-noise than the coadd, so the aggregate rate understates the confirmable fraction, and its complement is an upper bound on contamination rather than a measurement of it. The eccentricity sample of Section~\ref{sec:res-ecc} refits up to sixteen visits per system instead. The confirmed systems change velocity, as expected for unresolved binaries (Figure~\ref{fig:dvmax}), with a median primary velocity change of $10.2~\kms$ for the $8{,}347$ systems with at least three epochs, compared with $0.3~\kms$ for the unconfirmed systems. Among these $8{,}347$ systems, $1{,}929$ ($23\%$) have primary velocity changes below $1~\kms$. Only $2.5\%$ of this group appear in published SB2 catalogs, compared with $26.5\%$ of the remaining systems. We do not release $q_{\rm dyn}$ for systems in this group. Many of them are probably single stars, since the false positives of the visit fit change velocity little. For the best-sampled systems the two components show anti-phase velocities, and the slope of the secondary velocity against the primary recovers the mass ratio (Figure~\ref{fig:rvcurves}).

We also measured the false-positive rate of the visit fit. We used $1{,}619$ dwarfs drawn at random from the searched sample, excluding the catalog, the $\sigma_v$ list, the benchmark, and the training set of the fit's single-star network. The fit prefers two components for $7.8\%$ of these stars. The rate is higher for single stars already flagged by the classifier. Among flagged benchmark controls with steady velocities, it is $21\%$ ($27$ of $126$). Most of these false positives have small velocity changes, so a motion cut removes most of them. Requiring at least three nights and a primary velocity change above $10~\kms$ reduces the rate to $0.8\%$ for random dwarfs and $7\%$ ($6$ of $82$) for flagged steady controls. This cut keeps $4{,}149$ confirmed catalog SB2, which make up the velocity-confirmed sample. If close to $40\%$ of the $17{,}990$ catalog SB2 with at least three nights are single stars, about $500$ of these $4{,}149$ systems may also be single. Both control tables are included in the release.

The mass ratios from the combined spectra and the visit velocities do not agree closely. We compare them for $5{,}755$ confirmed systems with at least three nights, a primary velocity change above $1~\kms$, and $q_{\rm dyn}$ inside the allowed range of $0.1$ to $1.5$. We flag values within $0.05$ of either limit, which account for $8.9\%$ of the values. Among the unflagged values, $17\%$ are above $1$. These cannot come from a main-sequence pair and indicate either that the component labels exchange between visits or that an amplitude is poorly measured. The median $q_{\rm dyn}$ is $0.84$, while the combined-spectrum median for the same systems is $0.93$. Their Spearman correlation is $0.28$. We also calculate a Wilson mass ratio $q_{\rm W}$ from the untied component velocities of each visit. For the $3{,}085$ systems with $0.05<q_{\rm W}<3$, its Spearman correlation with $q_{\rm dyn}$ is $0.67$, compared with $0.33$ for the combined-spectrum value. We therefore use $q_{\rm dyn}$ and $q_{\rm W}$ as the mass-ratio measurements where they are available. Both use the same visits, so they are not independent.

Second, the visit velocities identify systems the coadd misses. We find $3{,}943$ dwarfs outside the SB2 catalog with a DR19 pipeline velocity that changes by more than $10~\kms$ across three or more visits, making them candidates for the SB1 class the combined-spectrum search cannot flag. We exclude visit velocities with $|v|>400~\kms$, which come from failed pipeline fits. We also exclude changes above $480~\kms$, approximately the largest expected for a main-sequence pair. We have not fitted the visits of these systems, so some may be double-lined. The same cut is passed by at most $0.9\%$ of random single dwarfs. They are released in the multi-epoch supplement, marked by its \texttt{sb1} flag. This search runs only on the $\sigma_v$-triaged multiply-visited dwarfs, not on the full $238{,}205$, so its count should be compared with the $663$ SB1 of \citetalias{2018MNRAS.476..528E} rather than scaling with the fifteen-fold larger SB2 catalog. And $3{,}031$ of the confirmed multi-epoch SB2 have three or more nights spanning more than $20~\kms$ in primary velocity, and $1{,}977$ of these were observed on at least four nights. Of these $3{,}031$ systems, $1{,}878$ are absent from both the APOGEE SB2 catalog of \citet{2021AJ....162..184K} and the LAMOST-MRS SB2 catalogs of \citet{2022MNRAS.517..356K,2024MNRAS.527..521K}. For $467$ of the $3{,}031$ systems, the component separation changes by less than $5~\kms$ across visits. Of these, $437$ have $q>0.9$. Only $23$ of the $467$ systems appear in published SB2 catalogs, compared with $38\%$ of the full group of $3{,}031$. In these cases, the component labels can exchange between visits. Excluding them leaves $2{,}564$ systems.

\begin{figure*}
\centering
\includegraphics[width=\textwidth]{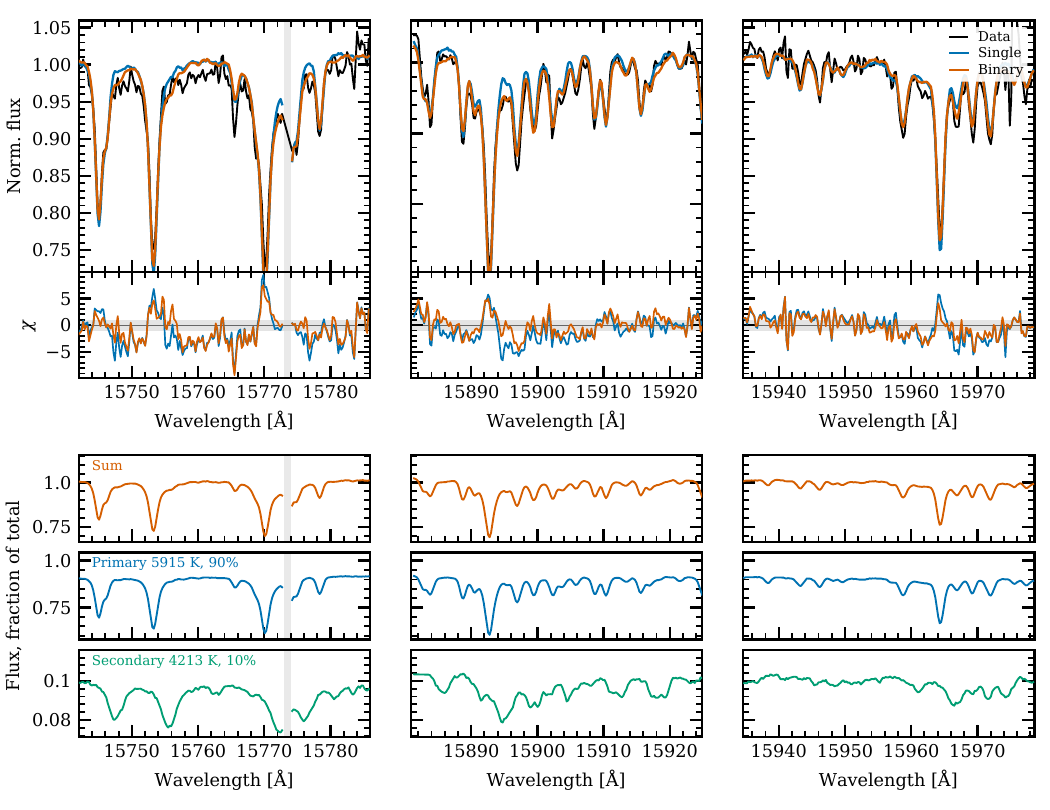}
\caption{The compact hierarchical triple LAMOST J042901.09+254144.2 (\Gaia\ DR3 151001777293928576), which our classifier flags from its combined spectrum, in the layout of Figure~\ref{fig:sb2fit}. We show our best fit to this spectrum, which gives $q=0.62$ and component velocities of $-0.2$ and $+44.5~\kms$. The primary has a temperature of $5{,}915$~K. The secondary is at $4{,}213$~K, contributes $10\%$ of the $H$-band flux, and has broadening close to the fit bound of $45~\kms$, above the $30~\kms$ limit of the network. The improvement is modest, with $\dchi=2{,}641$ and $\fimp=0.134$. This fit only narrowly passes the recalibrated gate. Its scaled value is $\fimp/0.90=0.149$, compared with the required rung of $0.14$ for this star. The catalog entry comes from the first census fit, which settled in a shallower minimum. That fit gives $q=0.92$, component velocities of $-0.56$ and $+4.65~\kms$, $\dchi=1{,}965$, and $\fimp=0.159$.\label{fig:refstar}}
\end{figure*}

One flagged system has an independent published orbit. \citet{2025arXiv250421287K} studied \Gaia\ DR3 151001777293928576, a contact eclipsing binary with a brighter third star on a $941$-day orbit, for which the archival APOGEE velocity is $-45.75~\kms$, while our combined-spectrum entry reads $-0.56$ and $+4.65~\kms$. The two do not disagree, since the velocities we recover from the coadd are splittings about the pipeline rest frame, rather than absolute velocities. We consider this a marginal detection. The catalog entry comes from the first census fit, which settled in a shallower minimum. Our best fit to the spectrum (Figure~\ref{fig:refstar}) gives a faint secondary with broad, shifted lines and only narrowly passes the acceptance gate. The two-component model cannot describe the very broad lines of a contact pair, and the fitted secondary reaches the fit bounds. So we do not identify this secondary with the inner binary found by \citet{2025arXiv250421287K}. Its single usable visit in the multi-epoch supplement carries no orbital information. Our fit to this visit gives $-45.88~\kms$, compared with $-45.75~\kms$ in both DR17 and DR19. For this star, the supplement gives the pipeline velocity.

\subsection{Eccentricities of the close twins}\label{sec:res-ecc}

The per-visit component velocities support one more population measurement. At wide separations, twin binaries ($q>0.95$) are far more eccentric than their non-twin peers, an excess that was measured from \Gaia\ astrometry \citep{2022MNRAS.512.3383H} and has been read as evidence that twins form in circumbinary disks at close separation and are widened afterward \citep{2022ApJ...933L..32H}. That measurement concerns separations of hundreds to thousands of au. The eccentricities of the close twins themselves, the regime in which the formation picture places them, have not been measured, and our SB2 sample reaches exactly that regime, since the periods we probe below correspond to separations of roughly $0.08$ to $1.2$~au, some three orders of magnitude inside the wide-twin measurement.

For the multi-epoch SB2 with eight or more visits we refit up to sixteen visits per system, selecting those with the highest signal-to-noise, and read the eccentricity from the shape of the two velocity curves. For each system we form the likelihood of its per-visit velocities as a function of eccentricity, with the other orbital elements integrated over and the velocity amplitude and systemic velocity solved in closed form (Appendix~\ref{app:orbit}). This gives one curve per system, the eccentricity likelihood, which does not yet involve the population. The population enters through a single index $\alpha$, with $f(e)\propto e^{\alpha}$, and we give the twins and the non-twins each their own $\alpha$ and compare the two. We keep the $520$ systems with median periods between $6$ and $400$~days, requiring confirmation by the visit fit and a constrained orbit.

\begin{figure}
\centering
\includegraphics[width=\columnwidth]{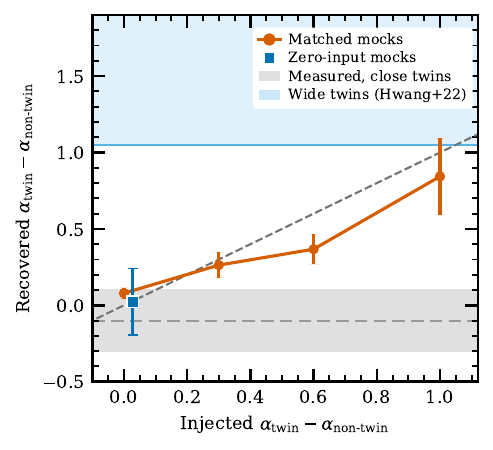}
\caption{Injection test of the eccentricity-index difference between the close twins and the matched non-twins. Mock populations with a known difference (orange) are recovered with a slope of $0.74$, below the one-to-one line shown by the dashes, and mocks with a zero difference at the real sampling (blue, offset for clarity) scatter around zero. The measured value, $-0.10\pm0.21$ (gray band), is consistent with zero, while the difference of the wide twins \citep[blue band,][]{2022ApJ...933L..32H} is of order $+1$ and would have been recovered.}\label{fig:eccalpha}
\end{figure}

The recovered index depends on how well each orbit is measured. A low-amplitude or sparsely sampled orbit spends most of its phase near apocenter, where the velocities move little, and it reads as more eccentric than it is. The velocity amplitudes are similar for twins and non-twins, with medians of $15.4$ and $15.0$~km/s, respectively. We still reweight the non-twins to match the twins in velocity amplitude and epoch count. This gives the two samples the same sampling distribution. We keep $309$ systems with primary amplitudes above $12$~km/s, including $131$ twins. On mock populations built at the real cadence the estimator carries a known difference through with a slope of $0.74$, so a true difference of $+1$ is recovered as about $+0.8$ (Figure~\ref{fig:eccalpha}), and putting the same index into both samples with the real twin and non-twin sampling returns $+0.02\pm0.05$, so the matching adds no difference of its own.

On the real systems the twins and non-twins have similar indices, with $\alpha_{\rm twin}=-0.20$ and $\alpha_{\rm non\text{-}twin}=-0.12$, and after the injection-test residual is removed the difference is $\alpha_{\rm twin}-\alpha_{\rm non\text{-}twin}=-0.10\pm0.21$, where the error combines a bootstrap over systems with the scatter of the injection test and of the matching choices. The value moves by at most $0.16$ when we match on period instead of amplitude, drop the systems below the $15$-day tidal circularization boundary \citep{2005ApJ...620..970M}, match on observed quantities alone, or keep only the systems both trained detectors flag, so the choices do not drive it. In mean eccentricity the two samples sit at $0.42$ and $0.45$, both inflated by the same apocenter effect, so only the difference should be read. The difference is $0.5$ standard deviations from zero, so the sample does not separate the two classes.

The two classes are therefore close in eccentricity, with a slight preference for the twins to be the more circular, of the sign that circularization of the more processed population would leave. An excess of the wide-binary sign is what the data exclude, since a true positive $\Delta\alpha$ above about $+0.75$, after accounting for the recovery slope, falls outside two standard deviations for every choice above, while the wide twins at $400$ to $1000$~au carry an index near $2$ against about $1$ for their non-twins \citep{2022ApJ...933L..32H}, a difference of order $+1$ that Figure~\ref{fig:eccalpha} shows our mocks would recover. We repeated the tests using the dynamical mass ratio to define twins. Only $22$ of the original $119$ twins retain that label. The new selection contains $45$ twins and gives a difference of $-0.02\pm0.32$. An excess of order $+1$ is still excluded, although less strongly. Whatever makes the wide twins eccentric is therefore not present at these separations, which supports the picture in which the excess is acquired during the widening rather than at birth. Appendix~\ref{app:orbit} gives the estimator, the tests, and the twin-specific caveats.

\subsection{The blind-agent test}\label{sec:res-blind}

The Skill records the operating decisions. A separate question is whether an agent could recover one on its own. We test this by removing one decision at a time from the Skill and handing a fresh agent the tools and the ablated version, with no access to the published answer, to see whether it rediscovers what was removed. We try two decisions that behave very differently.

Removing the self-consistent normalization (decision~1) normalizes the model and data with different operators, so every fit degrades. The median single-star $\chi^2$ over the controls rises from $2.1\times10^{4}$ to $6.2\times10^{4}$ for a few-percent continuum mismatch and to $2.4\times10^{5}$ for a larger one, a jump of a factor of three to more than ten that is plain in the controls alone, with no external label. The control false-positive rate, by contrast, does {\it not} rise: it falls, from 12\% to 3\% on the ablation's 250-star control subset (drawn at random from the benchmark controls) at the fixed acceptance, because the shared continuum error inflates the single-star and binary fits together and largely cancels in $\dchi$.

The signal the agent has to work with is therefore the absolute fit quality, and it is enough to act on. In a demonstration run, an agent given the high $\chi^2$ re-derived the common normalization. Because that signal is deterministic, we report it directly rather than as a success rate over many runs. The demonstration is a single run under a single model, and the two decisions below are the only ones we ablated, so the pattern is an illustration rather than a rate.

Removing the luminosity weighting (decision~8) behaves the opposite way. It biases the recovered mass ratio low, by up to $0.2$ at the equal-mass end, but leaves the fit quality and the acceptance unchanged, so nothing internal to the fit marks the change. Without an external mass-ratio reference the agent has no signal to act on, and the decision goes unrecovered.

\section{Discussion}\label{sec:disc}

\subsection{Comparison with \citetalias{2018MNRAS.476..528E}}\label{sec:disc-eb}

The agent reproduces the \citetalias{2018MNRAS.476..528E} decomposition on a new data release. At the detection stage the two agree closely (Section~\ref{sec:res-apogee}). The detection rate and the color--magnitude over-luminosity are comparable, and the median mass ratio over the common $0.2$--$0.95$ interval agrees ($0.80$ against $0.83$), at a benchmark-controlled false-positive rate (Section~\ref{sec:res-bench}). The DR19 catalog is about fifteen times larger in count, against a twelvefold growth of the searched dwarf sample from DR13 to DR19 ($238{,}205$ versus $20{,}142$). The modest excess over the survey growth may come from the recalibrated classifier's higher benchmark recovery at the catalog operating point ($74.4\%$ against about $55\%$ for the reference at the same $8.1\%$ false-positive rate; Section~\ref{sec:res-bench}), as well as contamination, since the two catalogs do not have matched purity.
The color--magnitude over-luminosity (Section~\ref{sec:res-apogee}) is a consistency check on a shared forward model, not an independent validation, because two selection-limited recovered distributions from the same model are expected to agree. The genuine external anchor is the \Gaia\ enrichment, the higher astrometric noise and non-single-star fraction of the flagged systems, drawn from data the decomposition never sees. The catalog also probes a different population from the radial-velocity-variability samples that dominate recent APOGEE multiplicity work \citep{2021AJ....162..184K}, which register a system only when its components are velocity-separated, whereas the combined-spectrum decomposition detects the two-temperature composite even at zero offset. The $11\%$ overlap (Section~\ref{sec:res-apogee}) is the expected consequence of that difference, not a purity measurement, all the more so as \citet{2021AJ....162..184K} work in the earlier DR16/17 reductions while our search is on DR19.

\subsection{Limitations}\label{sec:disc-limits}

Four inputs set limits we have not removed. The component temperatures and mass ratios come from a fixed $4$~Gyr solar-scaled isochrone, applied to a sample spanning disk, bulge, and halo, so the released $q$ and $T_{\rm eff,2}$ carry an age systematic we have not propagated, and the $q>0.95$ twin definition of Section~\ref{sec:res-ecc} rests on it. The $11{,}424$ searched primaries at the $4200$~K model floor are fit with labels the grid cannot reach, and we mark the $3{,}773$ that are included in the catalog with the \texttt{teff\_floor} flag in the release so they can be cut. The systems with eight or more visits come from the higher-cadence APOGEE fields, which include cluster and calibration pointings and are not a random draw from the survey, and we have not checked their cluster membership. Finally, \citetalias{2018MNRAS.476..528E} do not publish a control false-positive rate in a form matched to ours, so the comparison of our $17.4\%$ flagged fraction with their $13.1\%$ is between operating points of unknown relative purity.

The catalog is a set of spectroscopic detections, not a corrected multiplicity census. We report what the classifier flags and do not convert it into a completeness-corrected multiplicity fraction. That inference requires the two-dimensional $(q,\teff)$ selection function characterized in Section~\ref{sec:res-catalog} and Section~\ref{sec:data-valid}, and we leave it to future work.

The current tool set implements the single-versus-binary decomposition and its multi-epoch confirmation. In DR13, two higher-order classes were also fit from the 20{,}142 dwarfs \citepalias{2018MNRAS.476..528E}, $663$ single-lined (SB1) velocity-variable binaries and $114$ triple-lined (SB3) systems. Using the DR19 pipeline velocities, we identify $3{,}943$ candidates for the first class (Section~\ref{sec:res-multiepoch}). We do not match the second. We built the three-component extension and attempted the census, and it does not work on the DR19 products. A faint third star changes the spectrum by a few percent, and the single-star model trained on the public DR19 labels misses real spectra at a comparable level, so we cannot separate the two star by star. The limit sits in the training labels rather than in the spectra, and the same label noise is one reason the eccentricity comparison of Section~\ref{sec:res-ecc} uses the twin against non-twin differential rather than per-system eccentricities. We plan to improve the training labels next, and with them the triple census and the per-system orbits. The $3{,}031$ systems with primary velocity spans above $20~\kms$ are candidates for orbit fitting with the sampler in Appendix~\ref{app:orbit}. Meanwhile our pipeline either catches a triple as an SB2, when two of its components dominate the light, or rejects it at the acceptance gate, when the third component's light dilutes the fit improvement below the threshold.

\subsection{Single-star model quality and survey transfer}\label{sec:disc-instrument}

How well the whole system works on a survey is set by the quality of one component, the single-star spectral model. The decomposition flags a binary only when a genuine single star fits the single-star model down to the noise, so a model that reproduces real spectra poorly makes almost every star look faintly composite and blunts the single-versus-binary decision.

\citetalias{2018MNRAS.476..528E} built this model data-driven, training a network on a curated set of real APOGEE spectra. Reproducing that route from public ingredients recovers only $63.0\%$ of the benchmark SB2 at a common $13.3\%$ control false-positive rate, well short of the $75.6\%$ reached by the \citetalias{2018MNRAS.476..528E} network run as a best-case reference on the same benchmark (Section~\ref{sec:res-bench}), even though the two fit the SB2 spectra to the same residual ($\sim$0.83\%). This open reproduction is the first-step network of the training chain in Section~\ref{sec:analysis-apogee}. The gap is in single-versus-binary discrimination rather than fit accuracy, and it points to the tacit curation the open reproduction does not carry. \citetalias{2018MNRAS.476..528E} vetted their training sample by eye and censored contaminated or mislabeled stars before the network ever saw them. Label quality plausibly contributes as well, because the public DR19 labels used as training targets may carry more scatter than those of earlier releases such as DR13--14, and a network trained on noisier labels learns a fuzzier label-to-spectrum map. A distillation diagnostic locates the cause. When the same architecture is trained instead on noise-free, label-consistent spectra generated by the \citetalias{2018MNRAS.476..528E} network itself, it recovers $74.3\%$, so the gap lives in the training target rather than the network.

The training recipe of Section~\ref{sec:analysis-apogee} follows from this diagnosis. An ablation over the ingredients shows what each step buys, with every recovery quoted at its own control false-positive rate. Training on Kurucz synthetic models \citep{2003IAUS..210P.A20C} alone recovers $24\%$. Real spectra paired with the survey pipeline's labels, which come from a different code than the network, reach $54$--$63\%$. Refitting those labels with the network itself makes them self-consistent and yields the catalog model, which recovers $69.5\%$ at the original \citetalias{2018MNRAS.476..528E} acceptance thresholds. Those thresholds were tuned for the \citetalias{2018MNRAS.476..528E} network and leave ours at a $6.9\%$ false-positive rate, short of the $8.1\%$ we allow, so recalibrating them to the $8.1\%$ target (decision~5) moves the same network to the catalog's $74.4\%$, above the best-case reference at every matched false-positive rate (Section~\ref{sec:res-bench}). This self-consistency between the labels and the model is what gave the \citetalias{2018MNRAS.476..528E} network its edge, because their training labels were their own model's fits.

This single-star model is also the only part of the system tied to a specific instrument, through its line list, its spectral resolution, and the stars it was trained on, so it is the only part that does not transfer unchanged, and the recovery gap at equal fit quality above is what that costs. Moving to a new survey therefore means rebuilding this one model. The other seven servers and the whole Skill carry over unchanged, and the data server is a straight swap (Table~\ref{tab:transfer}). An empirical demonstration on a second survey is left to future work.

\subsection{What kind of know-how, and what an agent can recover}\label{sec:disc-knowhow}

The operating know-how behind a method like this falls into three categories that behave differently under transfer:
\begin{itemize}
\item {\it Codified} know-how is the equations and the acceptance thresholds: it is fully portable and is what a paper already communicates.
\item {\it Procedural} know-how is the order of operations and the decisions of Section~\ref{sec:analysis-skill}, the normalization convention, the masking, the multi-start, the calibration recipe: it is portable once written down, which is what the Skill does, but is normally left implicit.
\item {\it Calibrated} know-how is the survey-specific numbers, chiefly the single-star model that reproduces one instrument's spectra: it does not transfer at all and must be rebuilt, as discussed in Section~\ref{sec:disc-instrument}.
\end{itemize}
What an agent can recover of this on its own falls along the same lines. The blind-agent test (Section~\ref{sec:res-blind}) probes the boundary between the procedural and calibrated categories. An agent recovers procedural know-how that leaves a measurable trace, such as the normalization whose absence inflates the single-star fit $\chi^2$ on every control, and cannot recover calibrated know-how that needs an external anchor, such as the luminosity weighting whose absence is invisible without a mass-ratio reference.

The procedural layer is itself uneven. Some of it is published but scattered across the source paper, the label pipeline, and the reduction, so an agent that reads across all of it could in principle reassemble it, while some is genuinely tacit, like the priors for iterative clean-star selection and the judgment that recognizes an overfit. Automating the survey-specific calibration as a constrained, verify-gated loop with recovery on a labeled benchmark as the objective is the natural next capability, and the direction in which the per-survey human step can shrink.

This taxonomy is not special to binary detection. The same split, portable procedure versus instrument-specific calibration, appears wherever a data-analysis method is moved between facilities, and the practice of packaging the procedural layer as machine-runnable tools plus a written Skill is the same one now emerging around scientific agents in other domains \citep{2023arXiv230405376B, 2023Natur.624..570B, 2024arXiv240806292L, 2026OJAp....9.1879T}. In an era in which agents increasingly execute analyses, we expect the procedural layer, long left implicit, to be written down and published as a first-class artifact.


\section{Conclusion}\label{sec:conc}

Over the $238{,}205$ main-sequence dwarfs of APOGEE DR19, the \citetalias{2018MNRAS.476..528E} decomposition identifies $41{,}466$ double-lined spectroscopic binary candidates, $17.4\%$ of the searched sample and about fifteen times the $2{,}645$ dwarf SB2 reported by \citetalias{2018MNRAS.476..528E}, although the two samples do not have matched purity, against a twelvefold growth of the searched sample since DR13. The classifier reaches this at a validated $8.1\%$ control false-positive rate, with a benchmark recovery that exceeds the \citetalias{2018MNRAS.476..528E} network at every matched false-positive rate. The catalog is characterized by:
\begin{itemize}
\item a composite-model mass-ratio distribution that rises toward equal masses, with median $q=0.91$ ($0.80$ over the $0.2$--$0.95$ range where the decomposition is most sensitive), including the near-equal-mass twin excess;
\item component temperatures, tied through the isochrone, that pair Sun-like primaries (median $\teff\approx5{,}505$~K) with cooler K-type secondaries (median $\teff\approx4{,}590$~K);
\item independent \Gaia\ corroboration, with $77\%$ of the systems that have a reliable parallax sitting above the single-star main sequence, elevated astrometric noise ($49\%$ at RUWE$>1.4$ against $19\%$ of unflagged dwarfs), and more non-single-star solutions ($16\%$ against $5\%$);
\item a multi-epoch supplement that prefers two stars for $46\%$ of the SB2 with multiple visits and $94\%$ of those also identified as double-lined in published surveys, confirms velocity changes in $4{,}149$ systems, adds $3{,}943$ velocity variables outside the catalog, and identifies $3{,}031$ systems with primary velocity spans above $20~\kms$;
\item no significant difference in eccentricity between the close, well-sampled twins and matched non-twins, $-0.10\pm0.21$ in the population index, which excludes at these separations a twin eccentricity excess of the sign found at wide separations by \citet{2022ApJ...933L..32H}. The result also holds when twins are defined using the dynamical mass ratio, giving a difference of $-0.02\pm0.32$.
\end{itemize}
This is the largest sample of double-lined spectroscopic binary candidates assembled from APOGEE, released with per-system composite-model and, where the visits provide enough information, dynamical mass ratios, component velocities, and \Gaia\ and multi-epoch flags for higher-purity selection. At the control false-positive rate it carries of order $16{,}000$ falsely flagged singles, so users who need purity rather than completeness should cut on those flags.

The catalog was produced without re-deriving the method, by publishing its operating know-how rather than only its equations, packaged as agent-callable MCP tool servers and a written Skill of the decisions that make it work. An ablation of two decisions illustrates that this know-how is what carries the method, with a fresh agent recovering a removed decision when its absence leaves a measurable trace in the fit.

Almost all of this transfers to a new survey unchanged. Only the single-star spectral model, which carries one instrument's flux calibration, must be rebuilt (Table~\ref{tab:transfer}). Automating that last step as a verify-gated loop, with recovery on a labeled benchmark as the objective, is the clearest path to a binary classifier that opens on any survey with no human in the per-survey path, from the other SDSS-V spectroscopic programs \citep{2017arXiv171103234K} to the optical surveys that share the composite-fitting logic but not the wavelength range, LAMOST, WEAVE, 4MOST, and the \Gaia\ RVS spectra \citep{2012RAA....12.1197C, 2024MNRAS.530.2688J, 2019Msngr.175....3D, 2023A&A...674A..29R}.

More broadly, as language-model agents begin to carry out analyses, we expect the operating know-how of a method, long left implicit in code and expert practice, to become a research product in its own right. Publishing a method for an agent to run will increasingly mean publishing the decisions that make it work, alongside the equations.

\section*{Data Availability}
The MCP tool servers, the written Skill (whose operating decisions are enumerated in Section~\ref{sec:analysis-skill}), the DR19 SB2 catalog with its multi-epoch supplement and column descriptions, the orbit posterior summaries of the systems with eight or more visits, which carry the per-system eccentricity likelihoods under the uniform prior and are therefore unaffected by the population normalization of Appendix~\ref{app:orbit}, and the benchmark star list are released at \url{https://github.com/seratsaad/agent4binary}. The APOGEE DR19 spectra are public through the SDSS archive, and the catalog classifier is built entirely from public DR19 data, with the training manifest, the refit labels, and the trained network weights all released.

The catalog is produced by a deterministic science core, the single-versus-binary fit, the detection statistic, and the acceptance gate and multi-epoch confirmation behind it, and is reproducible from these servers under a fixed driver script. The language-model agent orchestrates and audits this core rather than replacing it, and where it runs, on the benchmark and the worked examples, it uses \texttt{gemini-2.5-flash}, so every number we report reads out of the deterministic core and does not depend on the choice of language model. The one component we cannot redistribute is the \citetalias{2018MNRAS.476..528E} reference network used for the benchmark comparison. Its \texttt{binspec} code and method are public, but its trained weights and curated training set are not ours to share, so the reference recovery numbers can be checked against our figures rather than regenerated from scratch.

\section*{Acknowledgments}

We thank Hans-Walter Rix and Kareem El-Badry for reading the manuscript and for their comments. We also thank the referee, Mikhail Kovalev, for the helpful suggestions that improved this paper. SMS is supported by the Distinguished University Fellowship awarded by The Ohio State University. YST is
supported by NSF under Grant AST-2406729 and by a Humboldt Research Award from the Alexander von Humboldt Foundation.

Funding for the Sloan Digital Sky Survey V has been provided by the Alfred P.\ Sloan Foundation, the Heising-Simons Foundation, the National Science Foundation, and the Participating Institutions. SDSS acknowledges support and resources from the Center for High-Performance Computing at the University of Utah. SDSS telescopes are located at Apache Point Observatory, funded by the Astrophysical Research Consortium and operated by New Mexico State University, and at Las Campanas Observatory, operated by the Carnegie Institution for Science. The SDSS web site is \url{www.sdss.org}.

This work has made use of data from the European Space Agency (ESA) mission {\it Gaia} (\url{https://www.cosmos.esa.int/gaia}), processed by the {\it Gaia} Data Processing and Analysis Consortium (DPAC, \url{https://www.cosmos.esa.int/web/gaia/}\allowbreak\url{dpac/consortium}). Funding for the DPAC has been provided by national institutions, in particular the institutions participating in the {\it Gaia} Multilateral Agreement.

This project has also benefited from using Claude Code for coding development and copy editing.

\facilities{Sloan (APOGEE), \Gaia}
\software{Astropy \citep{2013A&A...558A..33A, 2018AJ....156..123A, 2022ApJ...935..167A},
NumPy \citep{2020Natur.585..357H}, SciPy \citep{2020NaMet..17..261V},
Matplotlib \citep{2007CSE.....9...90H}, scikit-learn \citep{2011JMLR...12.2825P},
PyTorch \citep{2019arXiv191201703P}, LangGraph, the Model Context Protocol Python
SDK, and the \texttt{gemini-2.5-flash} language model}

\bibliographystyle{mnras}
\bibliography{refs}

\appendix

\section{The joint Keplerian sampler}\label{app:orbit}

For system $j$ with $N_j$ visits we use the component velocities from each visit's separate two-component fit, without imposing the momentum tie. We treat these as an unordered pair. For $q_j$, we use the dynamical mass ratio from the joint fit (Section~\ref{sec:res-multiepoch}). If this value is missing or at a bound, we use the combined-spectrum value. The two stars share one Keplerian orbit, described by the period $P$, eccentricity $e$, argument of periastron $\omega$, a reference phase $\phi$, the primary velocity semi-amplitude $K_1$, and the systemic velocity $v_{\rm sys}$. We advance the mean anomaly linearly in time, solve Kepler's equation for the eccentric anomaly by Newton iteration, and convert to the true anomaly $\nu_i$ in the standard way. The two line-of-sight velocities then share one orbital shape $g_i$,
\begin{equation}\label{eq:sb2kepler}
v_{{\rm rad},1,i}^{\rm mod} = v_{\rm sys} + K_1\,g_i , \qquad
v_{{\rm rad},2,i}^{\rm mod} = v_{\rm sys} - \frac{K_1}{q}\,g_i , \qquad
g_i \equiv \cos(\nu_i+\omega) + e\cos\omega ,
\end{equation}
where the primary semi-amplitude $K_1 = 2\pi a_1 \sin i / (P\sqrt{1-e^2})$ absorbs the inclination, and the secondary amplitude is fixed to $K_1/q$ by the dynamical mass ratio, so each epoch contributes two data points tied by one orbit. Radial velocities carry no plane-of-sky information, so the ascending node drops out, the inclination enters only through $K_1$, and the mass ratio $q=K_1/K_2$ stays free of it.

The nonlinear parameters carry wide priors, log-uniform in period from $0.5$ to $5{,}000$~days and uniform in $e$ on $[0,0.95]$ and in $\phi$ and $\omega$. At fixed $(P, e, \omega, \phi)$ the model of Equation~(\ref{eq:sb2kepler}) is linear in $(K_1, v_{\rm sys})$, which we solve in closed form and marginalize as a Gaussian. The archive does not release a velocity uncertainty per visit, so one per-epoch dispersion $s$ stands for all systems. Near $q=1$ the decomposition can exchange the two components between visits, so we sum the likelihood over both assignments at each epoch, which marginalizes the labels at fixed orbit.

For each system we then integrate over the nonlinear parameters by Monte Carlo and record the likelihood as a function of eccentricity alone, $\mathcal{L}_j(e)$, on a fixed grid of $36$ points, finer below $e=0.12$ to carry the weight a near-circular population places there, with $6\times10^{4}$ draws per grid point. This curve is the input to the population step and does not depend on the population index. There are $1{,}389$ confirmed systems with at least eight usable visits. Of these, $1{,}052$ have constrained orbits, and $520$ fall within the $6$ to $400$~day window.

\begin{figure}
\centering
\resizebox{0.85\textwidth}{!}{%
\begin{tikzpicture}[
    scale=1.3,
    shorten >=1pt,->,draw=black!70,
    neuron/.style={circle,minimum size=40,inner sep=0},
    param/.style={neuron,fill=cyan}, 
    inter/.style={neuron,fill=black!40}, 
    gauss/.style={neuron,fill=red}, 
    obs/.style={neuron,fill=yellow}, 
    box/.style={rectangle,draw=black,rounded corners,thick,inner sep=10pt},
    ebox/.style={rectangle,draw=black,dashed,rounded corners,thick,inner sep=8pt}
]

\node[param] (alpha0) at (-1.5,6.0) {$\alpha$};
\node[gauss] (palpha) at (0,6.0) {$e^{\alpha}$};

\node[param] (P)     at (-3,4) {$P$};
\node[param] (M0)    at (-1.5,4) {$\phi$};
\node[param] (omega) at (3,4) {$\omega$};
\node[param] (s)     at (4.5,4) {$s$};
\node[obs]   (q)     at (-3,-6) {$q$};

\node[gauss] (uP)  at (-3,2.5) {$\mathcal{U}$};
\node[gauss] (uM0) at (-1.5,2.5) {$\mathcal{U}$};
\node[gauss] (uom) at (3,2.5) {$\mathcal{U}$};
\node[gauss] (nq)  at (-1.5,-6) {$\mathcal{N}$};

\node[inter] (e)  at (0,2.5) {$e$};
\node[inter] (K1) at (1.5,2.5)
  {\parbox{1.2cm}{\centering $K_1$ \\ $v_{\rm sys}$}};
\node[inter] (R)  at (3,0.75) {$R$};

\node[inter] (nu) at (-2.25,0.75) {$\nu$};
\node[inter] (z)  at (-3.75,-0.75) {$z$};
\node[inter] (vmod) at (-0.75,-0.75)
  {\parbox{1.2cm}{\centering $v_1$ \\ $v_2$}};
\node[inter] (vrad) at (1.75,-0.75)
  {\parbox{1.4cm}{\centering $v_{{\rm rad},1}$ \\ $v_{{\rm rad},2}$}};
\node[gauss] (lik) at (1.75,-2.5) {$\mathcal{N}_{2}$};
\node[obs] (v1obs) at (0.25,-4) {$v_{{\rm rad},1}$};
\node[obs] (v2obs) at (3.25,-4) {$v_{{\rm rad},2}$};

\draw[->] (alpha0) -- (palpha);
\draw[->] (palpha) -- (e);
\draw[->] (P) -- (uP);
\draw[->] (uP) -- (nu);
\draw[->] (M0) -- (uM0);
\draw[->] (uM0) -- (nu);
\draw[->] (e) -- (nu);
\draw[->] (omega) -- (uom);
\draw[->] (uom) -- (R);
\draw[->] (e) -- (vmod);
\draw[->] (nu) -- (vmod);
\draw[->] (K1) -- (vmod);
\draw[->] (q) -- (nq);
\draw[->] (nq) -- (vmod);
\draw[->] (z) -- (vmod);
\draw[->] (vmod) -- (vrad);
\draw[->] (R) -- (vrad);
\draw[->] (vrad) -- (lik);
\draw[->] (s) to[out=270,in=20] (lik);
\draw[->] (lik) -- (v1obs);
\draw[->] (lik) -- (v2obs);

\node[ebox, fit=(nu)(z)(vmod)(vrad)(lik)(v1obs)(v2obs)] (epochbox) {};
\node[anchor=south west,font=\small] at ([xshift=-30pt,yshift=-40pt]epochbox.south east) {For each system};

\node[box, fit=(P)(M0)(omega)(s)(q)(uP)(uM0)(uom)(nq)(e)(K1)(R)(epochbox)] (bigbox) {};
\node[anchor=south west,font=\small] at ([xshift=20pt,yshift=70pt]bigbox.south west) {For each epoch};

\node[anchor=east] at ([xshift=-0.5cm]bigbox.west) {
    \begin{tikzpicture}[every node/.style={anchor=west}, x=1cm, y=0.6cm]
        \node[param,label=right:{Parameter}] at (3,6) {};
        \node[inter,label=right:{Intermediate Variables}] at (3,3) {};
        \node[gauss,label=right:{Distribution}] at (3,0) {};
        \node[obs,label=right:{Observed Variables}] at (3,-3) {};
    \end{tikzpicture}
};

\end{tikzpicture}
}
\caption{The hierarchical model behind the population comparison. Blue circles are model parameters, red are distributions, gray are intermediate variables, and yellow are observed, the per-visit radial velocities and the dynamical mass ratio $q$. The model draws a measurement uncertainty on $q$, while the fits hold $q$ at its measured value. The global index $\alpha$ sits above the per-system plate and ties the eccentricities together through the population prior $f(e)\propto e^{\alpha}$, estimated from the per-system eccentricity likelihoods.}\label{fig:orbitmodel}
\end{figure}
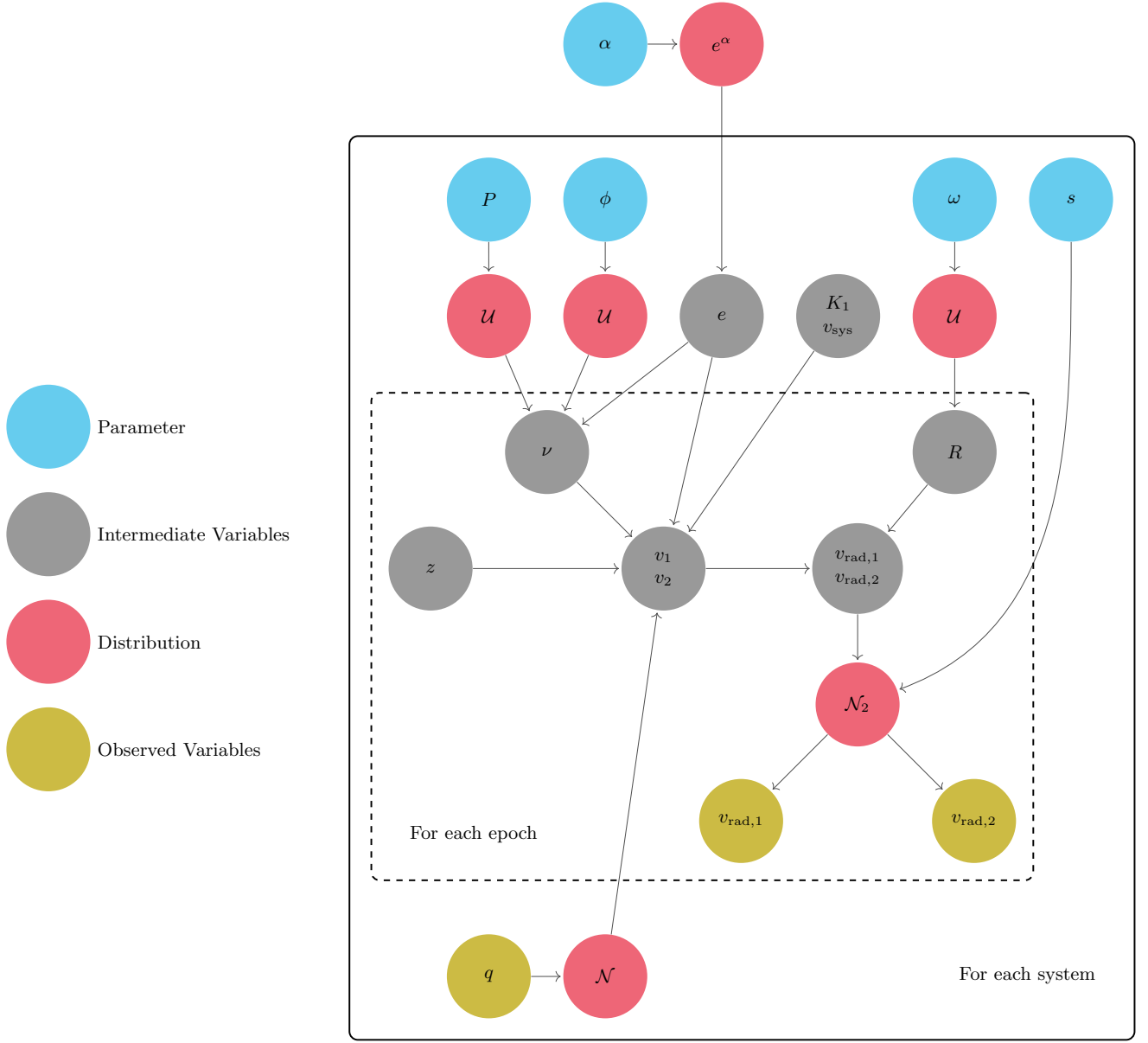

We model the population as $f(e\,|\,\alpha) \propto e^{\alpha}$ on $[0,e_{\rm max}]$ with $e_{\rm max}=0.95$, the same range as the per-system prior, normalized for any $\alpha>-1$, with $\alpha$ the global parameter at the top of Figure~\ref{fig:orbitmodel}. The value $\alpha=-1$ is the circular limit and sits at the edge of the allowed range, not inside it. Hierarchical inference of this kind on sparsely sampled APOGEE velocities was carried out by \citet{2020ApJ...895....2P}, and our step differs mainly in that we work from the per-system eccentricity likelihood rather than from posterior samples. The population log-likelihood folds each per-system curve against this prior by quadrature on the same grid,
\begin{equation}
\ln \mathcal{L}(\alpha) = \sum_{j} \ln\!\int_0^{e_{\rm max}} \mathcal{L}_j(e)\,
\frac{(1+\alpha)\,e^{\alpha}}{e_{\rm max}^{\,1+\alpha}}\,{\rm d}e ,
\end{equation}
and we maximize it over $\alpha$. We fit the twins and the non-twins separately, and before the fit we reweight the non-twins to the twins' joint distribution of velocity amplitude and epoch count, so the two samples carry the same sampling.

Two tests set how we read the result. We build mock populations of known index at the real cadence and amplitude distribution and pass them through the same steps. The recovered difference between two mock samples follows the input with a slope of $0.74$ and an offset of $+0.04$ (Figure~\ref{fig:eccalpha}), so a real difference is carried through at about three quarters of its original size. We also put one index into both mock samples while keeping the real twin and non-twin sampling. The recovered difference then stays at $+0.02\pm0.05$ over $24$ realizations at two input values, so the sampling difference between the two classes does not make a difference on its own. We still make no absolute claim for the data, since the archive does not release per-visit velocity errors, and a misstated dispersion moves the absolute index of both samples together, which is what the differential removes.

In the mocks, we randomly exchange the two velocities at each visit. The tests therefore include the ambiguity in the component labels. Three twin-specific effects sit outside these tests, and we note the direction of each. Line blending near conjunction acts on near-equal pairs alone and pushes $\alpha_{\rm twin}$ down, in the direction of the measured difference. The mass ratio is held at the dynamical value from the joint fit, without propagating its uncertainty, which has the largest effect near $q=1$. Contamination runs the other way, since the twins carry slightly more of it and noisy velocities read as eccentric, which would raise $\alpha_{\rm twin}$. None of these cancels in the differential, which is one reason we report a bound rather than a difference.

\end{document}